# New wine in old bottles: Quantum measurement – direct, indirect, weak – with some applications

Bengt E Y Svensson

*Theoretical High Energy Physics,*
*Department of Astronomy and Theoretical Physics,*
*Lund University,*
*Sölvegatan 14A,*
*SE-223 62 Lund, Sweden*

**Abstract**

The quantum theory of measurement has been with us since quantum mechanics (QM) was invented. It has recently been invigorated, partly due to the increasing interest in quantum information science. In this, partly pedagogical review, I attempt to give a self-contained overview of the basis of (non-relativistic) QM measurement theory expressed in density matrix formalism. I will not dwell on the applications in quantum information theory; it is well covered by several books in that field. The focus is instead on applications to the theory of weak measurement, as developed by Aharonov and Vaidman and their collaborators. Their development of weak measurement combined with what they call 'post-selection' – judiciously choosing not only the initial state of a system ('pre-selection') but also its final state – has received much attention recently. Not the least has it opened up new, fruitful experimental vistas, like novel approaches to amplification. But the approach has also attached to it some air of mystery. I will attempt to 'de-mystify' it by showing that (almost) all results can be derived in a straight-forward way from conventional QM. Among other things, I develop the formalism not only to first order but also to second order in the weak interaction responsible for the measurement. This also allows me to derive, more or less as a by-product, the master equation for the density matrix of an open system in interaction with an environment. One particular application I shall treat of the weak measurement is the so called Leggett-Garg inequalities, a k a 'Bell inequalities in time'. I also give an outline, even if rough, of some of the ingenious experiments that the work by Aharonov, Vaidman and collaborators has inspired.

If anything is magic in the weak measurement + post-selection approach, it is the interpretation of the so called weak value of an observable. Is it a *bona fide* property of the system considered? I have no answer to this question; I shall only exhibit the pros and cons of the proposed interpretation.

---

e-mail: Bengt_E_Y.Svensson@thep.lu.se

# I. Introduction

The measurement aspects of quantum mechanics (QM) are as old as QM itself. When von Neumann wrote his overview [1] of QM in 1932, he covered several of these aspects. The review volume [2] edited by Wheeler and Zurek tells the history up until ~1980. Since then, the interest has if anything increased, in particular in connection with the growth of quantum information science. Three books reflect the evolution of the field: the one by Braginsky and Khalil [3] from 1992, the one by Nielsen and Chuang [4] from 2000, and the more recent one by Wiseman and Milburn [5] from 2009. One important feature in the development over the last decades has been the increasing emphasis on what is call 'ancilla' (or 'indirect') measurement, in which the measurement process is modeled in some detail by considering how the system under study interacts with the measurement device.

Since the seminal paper [6] from 1988 by Aharonov, Albert and Vaidman, there has been another interesting development largely parallel to other ones. This is the field of weak measurement and weak values. Since it was started, its ideas have been skillfully advocated, in particular by Aharonov and Vaidman and their collaborators; see , *e.g.*, [7-9]. But it has also attracted attention by many others; a search on arXive.org for "weak measurement" gave (on Jan 9, 2012) some 15 hits for 2011.

Aharonov et al started from the well-known characteristics of QM which distinguishes it from classical physic: a QM measurement irrevocably disturbs the system measured. What, they asked, would happen if the interaction responsible for the measurement becomes very weak? True, the disturbance will decrease but so will also the information obtained in the measurement process. The revolutionary observation by Aharonov and collaborators is that this trade-off could be to the advantage of the non-disturbance aspect. There is new physics to be extracted from weak measurement, in particular when the weak measurement is combined with a judicious choice of the initial state ('pre-selection') and the final state ('post-selection') of the system under study.

This article provides an overview of the field of weak measurement without claiming originality of the results. The purpose is also to give a general pedagogical introduction to some more modern aspects of QM measurement theory. Prerequisites from readers will only be basic knowledge of the standard formulation of QM in Hilbert space as presented in any well-written text-book on QM, like [10].

To make the article self-contained, and to introduce the notations to be used, I begin anyhow with a summary of the standard QM theory of measurement. I treat both the direct (or projective) and the indirect (or ancilla) schemes. I do this in the density matrix formalism, which I also describe in some detail; other names for the density matrix are state matrix, density operator and statistical operator. Part of my motivation for doing so is that this formalism – arguably slightly more general than the pure-state, wave-function (or Hilbert-space state) formalism – is a straight-forward, though maybe somewhat clumsy, approach to reaching the QM results one wants, and this, again arguably, in a less error-prone way. In fact, the density matrix formalism provides an indispensable tool for any thorough-going analysis of the measurement process. Despite my preference for density matrices, in some of the more concrete problems I revert to the pure-state formalism. Still, I illustrate the advantage of the density matrix formalism by a side theme, not related to the main track of the article, *viz*., continuous measurement. In particular I derive, even if in a pedestrian way, the so called master equation for an open system

The approach to the measurement process taken here is not the most general one. An approach based on so called 'measurement operators' and on 'effects', alias POVMs (see, *e.g.*, [5] and my Appendix 1) is more general. In fact, the basic premises of that approach can be derived from the density matrix treatment of the ancilla method reviewed here. In Appendix 1, I expose these items.

My treatment of weak measurements will be focused on the scheme by Aharonov *et al* who combines it with 'post-selection'. I present the main arguments and derive expressions for the relevant quantities. In fact, I do this not only to linear order in the strength of the interaction but to second order.

Even without invoking weak measurements, it is of interest to study amplification. This is done by focusing on that sub-ensemble of the total measured sample that arises from chosing only those events that end up in a particular final state. I give some examples on how such an amplification set-up could be implemented.

If there ever were any 'magic' connected with the derivation and application of weak value, I hope my presentation will get rid of them. For example, nowhere in my exposé will there occur any negative probabilities or complex values of a number operator. In fact, what I present is nothing but conventional QM applied to some particular problems. In short, I hope that the article will provide a framework that aims to clear up some of the inadvertencies regarding the concept of weak value that, in my opinion, can be found in the literature both by theorists and experimentalists.

I choose to illustrate the fruitfulness of the weak measurement + post-selection approach by applying it to the so called Leggett-Garg inequalities, sometimes also called the 'Bell inequalities in time'.

One section of the article describes, even if in rough outlines only, some of the very ingenious experiments – admittedly subjectively chosen – that have exploited the new vistas opened up by the weak measurement + post-selection approach. To give a trailer of what they include: it is nowadays possible to measure the wave function of a particle directly, as well as to map the trajectories of the particles in a double-slit set-up.

A word on the (lack of) mathematical rigor in my treatment. I use mathematics the way physicists do, *i.e.*, essentially paying no attention to existence of the entities introduced, convergence of series expansions, *etc*. However, I think that, with appropriate restrictions and assumptions, the formalism can without too much labor be put on a sound mathematical base. I mention this because some of the criticism [12] of the approach by Aharonov *et al* has been the way they treat the mathematics. I do not subscribe to that objection. Even if their mathematics may sometimes seem to be a little cavalier, this criticism is minor and certainly does not wreck the whole edifice.

Up till the last but one section of this article, the arguments are strictly based on conventional QM and should not meet with any objections. If there are any open questions regarding the approach by Aharonov and collaborators, they concern the very interpretation of the formalism. In the penultimate chapter, I venture into this slightly controversial field. There, I attempt to further illuminate some of the basic premises that underlie the concept of a weak value, but also to raise some questions regarding its meaning and interpretation.

A final section gives a very short summary of the conclusions reached in the article.

Not covered in this article is the extension of conventional QM in terms of the so called the two-state vector formalism that Aharonov *et al* have proposed [7], partly as a further development of their weak measurement approach. These ideas definitely go beyond conventional QM.

Another active field of research, which is not covered herein, is the information-theoretic aspects of the QM measurement process. Here, I may refer the interested reader to, *e.g.*, [4,5].

*Remark.* During the final stage of writing this article, I became aware of the review article [48] by Kofman *et al*, with the same content as mine (and containing a much more extensive list of references than mine). The purpose and the basic tenor of that review is, however, different from mine: I want to give a general framework with emphasis on the basic principles, Kofman *et al* give a comprehensive review of all aspects of weak measurements.

# *II. The ideal (or projective) measurement scheme*

## *II A. The basic scheme*

All treatment of measurements in conventional quantum mechanics relies in one way or another on a standard treatment of the measurement process due to Born and von Neumann [1], with later extension by Lüders [13]. As it is described in most textbooks, it goes something like this.[1] The general situation is illustrated in figure 1.

The object to be measured, is a system $\mathbb{S}$ described by a (normalized) ket $|s>$ in a (finite-dimensional) Hilbert space $\mathcal{H}_{\mathbb{S}}$. This system could be an atom, an electron, a photon, or anything amenable to a quantum mechanical description. Suppose we are interested in measuring an observable $S$ on this system. The observable is described by a (Hermitian) operator[2] $\hat{S}$ and has a complete, orthonormal set of eigenstates $|s_i>$, $i = 1, 2, \ldots, d_S$, where $d_S$ is the dimension of the Hilbert space $\mathcal{H}_{\mathbb{S}}$. One may then expand $|s>$ in these eigenstates,

$$|s> = \sum_{i=1}^{d_S} c_i\,|s_i> = \sum_{i=1}^{d_S} |s_i><s_i|s> \ ,$$

where the complex coefficients $c_i$ obey

$$1 = \sum_{i=1}^{d_S} |c_i|^2 = \sum_{i=1}^{d_S} |<s_i|s>|^2,$$

expressing the fact that the state vector $|s>$ is normalized.

[1] The ideal or projective measurement scheme is sometimes also named after von Neumann. Since I will refer to a 'von Neumann protocol' later on in this article, I prefer not to attach von Neumann's name to the present scheme in order not to cause confusion.

So far it is mathematics. Now comes the physical interpretation:

(i) The possible result of measuring the observable $S$ on the system $\mathbb{S}$ is one of the eigenvalues $s_i$ of the operator $\hat{S}$, and nothing else.

(ii) Under the same conditions, the probability of obtaining a particular eigenvalue $s_i$ (assumed for simplicity to be non-degenerate) of the operator $\hat{S}$, given that the system $\mathbb{S}$ is in the state $|s>$, is

$$prob\,(s_i\,|\,|s>) = |c_i|^2 \,.$$

(iii) After such a measurement with the result $s_i$, the system $\mathbb{S}$ is described by the ket $|s_i>$. This is usually described as the 'reduction' (or 'collapse') of the state vector $|s>$ to $|s_i>$.

Moreover, I shall only consider non-destructive measurements, *i.e.*, measurements that do not destroy the system, even if they may leave it in a new state. For example, if the system is a photon, I require there still to be a photon after the measurement.

Let me elaborate somewhat on these basic premises.

Firstly, I want to reformulate the postulates in the more general framework, where states are described by density matrices, also called density operators or statistical operators. These are not vectors but operators on the Hilbert space $\mathcal{H}_{\mathbb{S}}$. For the simple case when the system $\mathbb{S}$ is described by a pure state in the form of a ket $|s>$, the corresponding density matrix – I will denote it by $\sigma$ – is simply the 'projector' $\Pi_s$ with the definition

$$\sigma := \Pi_s := |s><s| \qquad \text{(pure state)}\,.$$

In the more general case, when the system $\mathbb{S}$ is described by a (classical) statistical mixture, *i.e.*, an (incoherent) superposition of pure, normalized (but not necessarily orthogonal) states $|s^{(a)}>$, $a = 1, 2, \ldots,$ (say) $K$, each with its projector $\Pi_{s^{(a)}}$ and each with a probability $p_a$, the density matrix $\sigma$ representing the state of the system $\mathbb{S}$ is given by

$$\sigma_0 \ := \ \sum_{a=1}^{K} p_a \, \Pi_{s^{(a)}} \text{ (incoherent sum of pure states).}$$

---

[2] Here I distinguish the observable from the corresponding operator by putting a hat symbol on the letter designating the operator. I shall not be that careful later on and often use the same symbol for observable and operator. I hope this will not cause any confusion.

Here, I attached a subscript 0 to $\sigma$ in order to identify it as the initial state. We shall meet such density matrices in a slightly different context later on, when discussing the state of one part of a system composed of several subsystems.

One important property of a density matrix is expressed by its trace. The trace is defined for an arbitrary operator $\hat{A}$ as the sum of its diagonal elements:

$$Tr\,(\hat{A}) \ := \ \text{trace}\,(\hat{A}) \ := \ \sum_i <s_i|A|s_i>.$$

It follows that

$$Tr\,(\,\sigma_0\,) \ = \ Tr\,(\,\sum_{a=1}^{K} p_a \prod_{s^{(a)}}) \ = \ \sum_{a=1}^{K} p_a\, Tr\,(\,\prod_{s^{(a)}}\,) \ = \ \sum_{a=1}^{K} p_a \ = \ 1\,,$$

where the third equality follows from the normalization of the states $|s^{(a)}>$ and the last from the fact that all the probabilities $p_a$ must add up to 1. This trace property of any density matrix then expresses the normalization condition and is a very useful check when performing calculations.

With such an incoherent sum of pure states, the measurement rules (i) – (iii) above take a partly new form:

(i´) Same as (i).

(ii´) $prob\ (s_i \mid \sigma_0\,) \ = \ \sum_a p_a \ \ prob\ (s_i \,\big|\, |s^{(a)}>)\,.$

Before turning to the third ('collapse') condition in the density matrix formalism, I want to elaborate on this new expression for the probability. Using

$$prob\ (s_i \,\big|\, |s^{(a)}>) = \big|< s_i|s^{(a)}>\big|^2 = \ < \ s_i|s^{(a)} \ >< \ s^{(a)}|s_i > \ = \ < \ s_i|\Pi_{s^{(a)}}|s_i >\,,$$

one finds

$$prob\ (s_i \mid \sigma_0\,) \ = \ < \ s_i| \ \sum_a \ p_a\, \Pi_{s^{(a)}} \ |s_i > \ = \ < \ s_i|\,\sigma_0\,|s_i >.$$

It is standard to rewrite this simple expression in a slightly more complicated form, which also will be advantageous later on. Using the fact that the eigenstates are orthogonal, $< s_k\,|s_i > = \delta_{k,i}$ , one deduces that

$$prob\ (s_i \mid \sigma_0\ ) \ = \ < \ s_i|\,\sigma_0\,|s_i > \ = \ \sum_k < \ s_k\,|s_i >< \ s_i|\,\sigma_0\,|s_i >< \ s_i\ |s_k > =$$

$$= \ Tr\,(\prod_{s_i} \sigma_0 \prod_{s_i}) \ = \ Tr\,(\prod_{s_i} \sigma_0\,)\,.$$

Here the last equation follows from the easily proved invariance of the trace under cyclic permutations,

$$Tr\,(\hat{A}\ \hat{B}\ \hat{C}\,) = Tr\ \ (\hat{B}\ \ \hat{C}\,\hat{A})\,,$$

and from the fact that a projector $\Pi$ is idempotent, $\Pi \times \Pi = \Pi$.

As an interim summary, then, the second condition may be formulated as

(ii´) $prob\ (s_i \mid \sigma_0) = Tr\,(\prod_{s_i} \sigma_0)$.

The collapse condition may now be stated in two different forms, as 'conditional' or 'unconditional'. The conditional (or 'selective') form applies when one asks for the density matrix after the measurement with the a particular result $s_i$ and reads

(iii´) Measurement of *S*, with result $s_i$ , transforms the original density matrix $\sigma_0$ into

$$\sigma_0 \longrightarrow \frac{\Pi_{s_i}\ \sigma_0\ \Pi_{s_i}}{prob\ (s_i \mid \sigma_0)} =: \sigma_1(\ \mid s_i)$$

(Lüders' rule for a conditional density matrix),

with subscript 1 identifying entities after the measurement. This is the so called Lüders' rule [13] for how an ideal measurement transforms (or 'collapses' or 'up-dates') the initial density matrix; we will meet this rule in several disguises in the sequel.

Let me note in passing that the rules (ii´) and (iii´) in fact also apply in case the eigenvalue $s_i$ is degenerate, then with the projector onto a non-degenerate state replaced by the projector onto that subspace of the Hilbert space $\mathcal{H}_{\mathfrak{S}}$ which is spanned by the eigenvectors having this eigenvalue.

It is easy to check that the rule (iii´) entails the rule (iii) when the density matrix $\sigma_0$ represents a pure state. Moreover, one may convince oneself that the conditional density matrix, $\sigma_1(\ \mid s_i)$ , even if the initial state is a mixture, represents a pure state provided the eigenvalue $s_i$ is non-degenerate.

The 'unconditional' or 'non-selective' case occurs when the measurement is performed but when, for one reason or another, one does not register the outcomes. Then, one must average over the possible conditional density matrices, each with its probability, to get:

(iii´´) Measurement of *S*, but without registering the result, transforms the original density matrix $\sigma_0$ into

$$\sigma_0 \longrightarrow \sum_i\ prob\ (s_i \mid \sigma_0)\ \mathrm{x}\ \sigma_1(\mid s_i) = \sum_i\ (\prod_{s_i} \sigma_0 \prod_{s_i}) =: \sigma_1$$

(unconditional density matrix).

## *II B. Mean values. Interference*

It follows from the rules just given that the statistical mean value of the observable *S* in the state $\sigma_0$ is given by

$$< S >_0\ := \sum_i s_i\ prob\ (s_i \mid \sigma_0) = \sum_i s_i\ Tr\,(\prod_{s_i} \sigma_0) = Tr \sum_i\ (s_i \prod_{s_i} \sigma_0) =$$

$$= Tr\,(\hat{S}\ \sigma_0)$$

since the operator $\hat{S}$ has the spectral representation

$$\hat{S} = \sum_i s_i \ \Pi_{s_i} .$$

It is important to keep in mind that the mean value depends both on the operator $\hat{S}$ and on the state $\sigma_0$ .

The same final expression applies to any operator. As an application, consider the mean value of an arbitrary operator *A* , first in the initial state of figure 1:

$$< A >_0 \ := \ Tr\,(\hat{A}\,\sigma_0) \ = \ \sum_{i,j} < s_i |\ \hat{A}\ | s_j > < s_j | \sigma_1 | s_i >.$$

In this expression both diagonal and non-diagonal elements of the density matrix appear, which means that interference between the different matrix elements of $\hat{A}$ might occur: they add coherently. Compare this with the mean value of the same operator in the state after the measurement:

$$< A >_1 \ := \ Tr\,(\hat{A}\,\sigma_1) \ = \ \sum_{i,j} < s_i |\ \hat{A}\ | s_j > < s_j | \sigma_0 | s_i > \ =$$

$$= \sum_i < s_i |\ \hat{A}\ | s_i > < s_i | \sigma_0 | s_i >$$

since

$$< s_j | \sigma_1 | s_i > \ = \ \delta_{i\,,j} \ < s_i | \sigma_0 | s_i >.$$

So in this case there can be no interference: the sum is an incoherent sum over diagonal elements only.

## *II C. Post-selection and the ABL rule*

As another application of the projective measurement scheme, let me consider a situation in which two successive measurements are made on the same system; remember that I suppose all measurement on the system under study to be non-destructive. The situation is illustrated in figure 2.

Begin by preparing ('pre-selecting') the system $\mathbb{S}$ in an initial (pure, for simplicity) state $|s>$. Let this be followed by a measurement of the observable *S*, resulting in one of the eigenstates $|s_i>,\ i \ = \ 1, 2, \ldots, d_S,$ of the operator $\hat{S}$ with probability $prob\ (s_i \mid |s>) \ = = |< s_i | s >|^2$. Finally, subject the system to a second measurement, now of another observable *F*. One is interested in the case when this second measurement projects the system into a particular eigenstate $|f>$ with eigenvalue *f* of the corresponding operator. The joint probability for obtaining $|s_i>$ in the first measurement and $|f>$ in the second is then

$$prob\ (s_i, f \mid |s>) \ = \ prob\ (\,f\ \mid | s_i > ) \ \text{x} \ prob\ (s_i \mid |s> ) \ =$$

$$= \ |< f\, |\, s_i >|^2 \ \ \text{x} \ \ |< s_i\, | s >|^2$$

Moreover, taking into account that any of the $d_S$ states $|s_i >$ could be projected to $|f>$, the total probability $prob\ (f \mid |s>)$ of obtaining $f$ independently of the intermediate states $|s_i>$ is

$$prob\ (f \mid |s>) = \sum_i prob\ (s_i, f \mid |s>) = \sum_i |<f \mid s_i>|^2 \ \text{x}\ |<s_i|s>|^2$$

So far the argument is straight-forward. But let me now, with Aharonov, Bergmann and Lebowitz ('ABL') [14], so to speak turn the argument around and ask for the probability $prob\ (s_i \mid |f>, |s>)$ of finding a certain intermediate eigenvalue $s_i$, given that the post-selection obtains *f*. Using standard (Bayes') rules for handling probabilities, ABL deduced

$$prob\ (s_i \mid |f>, |s>) = \frac{prob\ (s_i, f \mid |s>)}{prob\ (f \mid |s>)} = \frac{|<f \mid s_i>|^2\ |<s_i|s>|^2}{\sum_j |<f \mid s_j>|^2\ |<s_j|s>|^2} \ .$$

This is the ABL-rule when the eigenvalue $s_i$ is non-degenerate. In case $s_i$ is a degenerate eigenvalue of the operator $\hat{S}$ this ABL rule reads

$$prob\ (s_i \mid |f>, |s>) = \frac{\left|<f \mid \Pi_{s_i} |s>\right|^2}{\sum_j \left|<f \mid \Pi_{s_j} |s>\right|^2} \ ,$$

where, as before, $\Pi_{s_i}$ is the projector onto the subspace corresponding to the eigenvalue $s_i$.

From this ABL rule one may now, *e. g.*, obtain a conditional mean value, $<S>_f$, of the observable *S*, i. e. conditioned on the outcome $|f>$ of the post-selection:

$$<\hat{S}>_f := \sum_i s_i\ prob\ (s_i \mid |f>, |s>) = \sum_i s_i \frac{|<f \mid s_i>|^2\ |<s_i|s>|^2}{\sum_j |<f \mid s_j>|^2\ |<s_j|s>|^2}$$

The ABL rule has been the focus of much further development, in particular by Aharonov and his collaborators. It has also stirred much controversy, relating, *e. g.*, to questions whether the rule can in some way be applied 'counterfactually', *i.e.*, even if the intermediate measurement of the observable *S* is not carried out. I will not go into anything of this, but refer the interested reader to some of the literature (see [7, 8] and references therein). I will, however, comment on the ABL rule in connection with my treatment below of so called weak measurements.

## *II D. Time evolution*

For completeness, let me quote the rules for time evolution in QM.

For a pure state $|s>$ at time $t_0$, one obtains the state $|s>_t$ at a later time *t* from the evolution equation

$$|s> \xrightarrow[\text{under } H_S]{\text{evolution}} |s>_t = \mathcal{U}\ |s> := exp\left(-i \int_{t_0}^{t} dt'\ H_S\right) |s>, \quad (\hbar = 1),$$

by time-integration of the Schrödinger equation with the Hamiltonian $H_{\mathbb{S}}$ . This translates immediately to the time evolution of a density matrix

$$\sigma_0 \xrightarrow[\;]{\substack{evolution \\ under\ H_{\mathbb{S}}}} \sigma_t := \mathcal{U}\, \sigma_0\, \mathcal{U}^\dagger ,$$

with $\mathcal{U}^\dagger$ the Hermitian conjugate of $\mathcal{U}$.

# *III. The indirect (or ancilla) measurement scheme*

## *III.A Modeling the measurement process*

The approach to measurement described above leaves the concept of 'measurement' un-analyzed. In particular, it takes no account whatsoever of how the measurement is performed, what kind of measurement apparatus is used, what distinguishes measurements from other possible types of interactions, etc.

In the indirect, or ancilla, scheme that I will now describe, one goes a few steps in the direction of describing the very measurement process. True, it is still very schematic. But at least it introduces a measurement device (or ancilla) – to be alternatively called a 'meter' or a 'pointer' – into the picture even if ever so schematically. It is the interaction of the meter with the system – called alternatively the object, or the probe or, for photons, the signal photon – that constitute the measurement: by reading off the meter one gets information as to the value of the system observable[3]. The scheme is described diagrammatically in figure 3.

The meter $\mathcal{M}$ will be modeled as a quantum device. It is assumed to have a Hilbert space $\mathcal{H}_{\mathcal{M}}$ with a complete, orthonormal set of basis states $|m_k>$, $k = 1, 2, ..., d_M$ , where $d_M$ is the dimension of $\mathcal{H}_{\mathcal{M}}$ . The operator in $\mathcal{H}_{\mathcal{M}}$ which has these states as eigenstates is denoted $M$. For reasons to become clearer later on, the observable $M$ is called the 'pointer variable', the states $|m_k>$ the 'pointer states'. The intrinsic Hamiltonian of $\mathcal{M}$ is denoted $H_{\mathcal{M}}$ (for simplicity, it will be assumed to vanish[4] in most cases I treat, but it is instructive to keep it when setting up the scheme), and projectors in $\mathcal{H}_{\mathcal{M}}$ will be denote $\Lambda_{m_k}$ , etc. The meter is assumed to be prepared in an initial pure state $|m^{(0)}>$ – not necessarily an eigenstate of the pointer variable $M$ (see below) – so that the meter initial density matrix is $\mu_0 = |m^{(0)}> < m^{(0)}|$ .

The object or system $\mathbb{S}$ – I sometimes also refer to it as the 'object-system' in order to distinguish it from the total system comprising the meter and the object-system – has its Hilbert space $\mathcal{H}_{\mathbb{S}}$ in the same way as in the projective measurement scheme of the previous section. It has a complete, orthonormal set of basis states $|s_i>$, $i = 1, 2, \ldots, d_S$ , with $d_S$ the

[3] In experimental situations, the 'meter' and the 'system' could even be properties of one and the same physical object – like momentum and polarization for a photon. See section VII for some examples.

[4] In case the intrinsic Hamiltonians do not vanish the technique is to use time-dependent operators in the Heisenberg representation. See*, e.g.,* [15].

dimension of the Hilbert space $\mathcal{H}_{\mathbb{S}}$. They are eigenstates of the operator $S$ in $\mathcal{H}_{\mathbb{S}}$ which corresponds to the observable $S$ to be measured (from now on, I consequently use the same notation for a quantum operator as for the corresponding observable). The system's intrinsic Hamiltonian is $H_{\mathbb{S}}$ (which, as with $H_{\mathcal{M}}$, I will assume to vanish); projectors in $\mathcal{H}_{\mathbb{S}}$ will be denoted $\Pi_{s_i}$, *etc*. The system is assumed to be initially prepared ('pre-selected') either in a pure state $|s> = \sum_{i=1}^{d_S} c_i\,|s_i>$ – in which case its density matrix is $\sigma_0 = |s><s|$ – or in a more general state described by an arbitrary $\sigma_0$ (but of course normalized so that $Tr\,\sigma_0 = 1$)

The total system $\mathcal{T}$ comprises the object-system $\mathbb{S}$ and the meter $\mathcal{M}$. Its Hilbert space is $\mathcal{H}_{\mathcal{T}} =$ $= \mathcal{H}_{\mathbb{S}} \otimes \mathcal{H}_{\mathcal{M}}$, where the symbol '$\otimes$' stands for the direct product. The initial state of the total system is $\tau_0 = \sigma_0 \otimes \mu_0$, i.e. the system and the meter are assumed to be initially uncorrelated (not entangled).The total Hamiltonian is $H_{\mathcal{T}} = H_{\mathbb{S}} + H_{\mathcal{M}} + H_{int}$. The interaction Hamiltonian $H_{int}$ does not vanish, and depends on the observable $S$ to be measured as well as on a pointer variable; see below for details. Moreover, the assumption of non-destructive measurement (as spelled out in in section *III.B.* below) – in the present case, with vanishing intrinsic Hamiltonians, this is the criterion for a so called quantum non-demolition, QND, measurement – implies that the commutator $[H_{int}, S] = 0$.

## *III.B. The pre-measurement*

The system and the meter are assumed to interact via a unitary time-evolution operator $\mathcal{U}$ in what is called a pre-measurement. This means that the total system $\mathcal{T}$ with its initial density matrix $\tau_0$ will evolve unitarily into $\tau_1$ :

$$\tau_0 = \sigma_0 \otimes \mu_0 \overset{\mathcal{U}}{\rightarrow} \tau_1 := \mathcal{U}\, \sigma_0 \otimes \mu_0\, \mathcal{U}^\dagger ,$$

where $\mathcal{U}^\dagger$ is the Hermitian conjugate of $\mathcal{U}$. If the Hamiltonian is known, the unitary operator $\mathcal{U}$ is given by $\mathcal{U} = exp\,(-\,i \int dt\, H_{\mathcal{T}})$ with $t$ = time (I use units so that $\hbar = 1$). To start with, I shall, however, not use this Hamiltonian expression but characterize $\mathcal{U}$ in another way.

To this end, note that for $\mathcal{U}$ to be a (pre-)*measurement* of $S$, $\mathcal{U}$ must have properties so that it distinguishes between the different states $|s_i>$. It is therefore assumed that an initial joint pure state $|s_i> \otimes |m^{(0)}>$ of the system and the meter – remember that I consider non-destructive measurements – is transformed by $\mathcal{U}$ into

$$|s_i> \otimes\, |m^{(0)}> \quad \overset{\mathcal{U}}{\rightarrow} \quad \mathcal{U}(\,|s_i> \otimes |m^{(0)}>\,) \;=:\; |s_i> \otimes |m^{(i)}> \;,\quad i = 1, 2, \ldots, d_S ,$$

where the meter states $|m^{(i)}>$ act as markers for the system state $|s_i>$ ; we will see in detail how this comes about later.

If this initial state is a superposition of eigenstates, $|\,s> = \sum_{i=1}^{d_S} c_i\ |s_i>$, this implies ($\mathcal{U}$ is a linear operator!)

$$|s> \otimes |m^{(0)}> \overset{\mathcal{U}}{\to} \mathcal{U}\,(|s> \otimes |m^{(0)}>) = \sum_{i=1}^{d_S} c_i\,|s_i> \otimes |m^{(i)}>.$$

The joint matrix, written out in full detail, evolves according to

$$\tau_0 = \sigma_0 \otimes \mu_0 \overset{\mathcal{U}}{\to} \tau_1 = \mathcal{U}\,\sigma_0 \otimes \mu_0\,\mathcal{U}^\dagger =$$

$$= \sum_{i,j} (\,|s_i> \otimes |m^{(i)}> \; <s_i|\sigma_0|s_j> \; <s_j| \otimes <m^{(j)}|\,) =$$

$$= \sum_{i,j} (\,|m^{(i)}> \Pi_{s_i}\,\sigma_0\,\Pi_{s_j} <m^{(j)}|\,),$$

where the last equation has introduced the projection operators $\Pi_{s_k} = |s_k> < s_k|$, $k = i, j$.

It is important to note that

- A system's pure eigenstate $|s_i>$ is left unchanged under this operation; among other things, the assumption of a non-destructive measurement is important here.
- One of the most important consequences of the pre-measurement is that the object-system's state becomes correlated (entangled) with the meter state: $\tau_1$ cannot be written as a product of one object state and one meter state.
- The meter state $|m^{(0)}>$ and $|m^{(i)}>$ are, in general, not eigenstates of the meter operator $M$, but superpositions of such eigenstates. In particular, the states $|m^{(0)}>$ and $|m^{(i)}>$, $i =$ $=1, 2, \ldots, d_S$, are normalized but in general not mutually orthogonal. Nor do they form a complete set in $\mathcal{H}_\mathcal{M}$. Indeed, the dimensions $d_S$ and $d_M$ of the respective Hilbert spaces $\mathcal{H}_\mathbb{S}$ and $\mathcal{H}_\mathcal{M}$ need not be equal.
- The operation $\mathcal{U}$ thus correlates the system state $|s_i>$ with the meter state $|m^{(i)}>$ but not necessarily in a unique way: to each $|s_i>$ there corresponds a definite $|m^{(i)}>$, different for different $|s_i>$, but there could be overlap between $|m^{(i)}>$ and $|m^{(j)}>$, expressed by $<m^{(i)}|m^{(j)}> \neq 0$, for $i \neq j$. An example we will see in more detail later on: $|m^{(i)}>$ could represent a Gaussian distribution for a continuous 'pointer variable' $q$ (the position of a pointer on a scale) so that $|m^{(i)}> \sim exp\left(-(q-g\,s_i)^2/4\,\Delta^2\right)$, where $g$ is a coupling constant and $\Delta$ the width of the Gaussian; for large enough $\Delta$ compared to $g\,s_i$, different wave functions (different $i$-values) will overlap. See further section IV.A below.

The rule for obtaining the separate states $\sigma_1$ for the system, and $\mu_1$ for the meter, after this pre-measurement, is to take the partial trace over the non-interesting degrees of freedom. In case we want the state $\sigma_1$ of the system, this means summing over the basis states $|m_k>$ for $\mathcal{H}_\mathcal{M}$, transforming an operator in $\mathcal{H}_\mathbb{S} \otimes \mathcal{H}_\mathcal{M}$ into one in $\mathcal{H}_\mathbb{S}$:

$$\sigma_0 \overset{\mathcal{U}}{\to} \sigma_1 := Tr_\mathcal{M}\,\tau_1 = \sum_k < m_k|\,\tau_1\,|m_k> =$$

$$= \sum_{k,i,j} < m_k|m^{(i)}> \; \Pi_{s_i}\,\sigma_0\,\Pi_{s_j} \; < m^{(j)}|m_k> =$$

$$= \sum_{i,j} (\,\Pi_{s_i}\,\sigma_0\,\Pi_{s_j} < m^{(j)}|m^{(i)}>\,).$$

The last equation follows from the completeness of the basis states $|m_k>$ in the Hilbert space $\mathcal{H}_{\mathcal{M}}$.

As is seen, were it not for a possible overlap between the meter states $|m^{(i)}>$ and $|m^{(j)}>$, the system density matrix $\sigma_1$ after the pre-measurement would have had diagonal elements only, i.e. would not have allowed any interference effects between different eigenvalues $s_i$. Another way to express this fact is to realize that, in case of no overlap between different states $|m^{(i)}>$, the indirect measurement scheme have the same consequences for the object-system as would a projective measurements have had:

$$< m^{(i)}|m^{(j)} > = 0 \text{ for } i \neq j \quad \Rightarrow \quad \text{(ancilla scheme)} \longrightarrow \text{(projective scheme)}$$

The general case, $< m^{(i)}|m^{(j)} > \neq 0$ for $i \neq j$, does allow for interference, a fact which will have interesting measureable effects as we will see in more detail in section V.

The same conclusion may be reached by considering the meter. Its state after the pre-measurement is

$$\mu_0 \overset{\mathcal{U}}{\rightarrow} \mu_1 := Tr_{\mathbb{S}}\, \tau_1 = \textstyle\sum_i < s_i\,|\,\tau_1\,|s_i > \; = \textstyle\sum_i |m^{(i)}> \;\; <s_i\,|\sigma_0\,|s_i> \; < m^{(i)}\,|\,,$$

with matrix elements

$$< m_k\,|\,\mu_1|\,m_l > \; = \textstyle\sum_i < m_k|m^{(i)}> \;\; < s_i\,|\,\sigma_0\,|s_i> \; < m^{(i)}\,|\,m_l>$$

expressed in terms of the wave functions $< m_k\,|\,m^{(i)} >$. From this result we again see clearly that a pre-measurement can distinguish effectively between different values $s_i$ of the system observable $S$ only if different meter wave functions $< m_k\,|m^{(i)} >$ do not overlap.

## *III.C. The read-out: Projective measurement on the meter*

So far, no real measurement has been performed in the sense of obtaining a record. The entangled system-meter is still in a quantum-mechanical superposition $\tau_1$. One needs a recording, a read-out of the meter, in order to obtain information that constitutes a real measurement.

Therefore, the next step in this indirect measurement scheme is to subject the meter, and only the meter, to a projective measurement of the pointer variable $M$. Reading off the meter means obtaining an eigenvalue $m_k$ of $M$ in a reaction that is symbolized by the projector $\Lambda_{m_k} := |m_k> < m_k|$ onto the corresponding subspace of the meter Hilbert space $\mathcal{H}_{\mathcal{M}}$. Since, as a result of the pre-measurement, the system is entangled with the meter, this will influence the system state too, and is therefore also a measurement of the object-system as will be evident shortly.

For the total density matrix, this projective measurement, complying to the Lüders' rule, implies

$$\tau_1 \xrightarrow{\Lambda_{m_k}} \tau_1(\,|m_k\,) := \frac{(\mathbb{1}_{\mathcal{S}} \otimes \Lambda_{m_k}\,)\ \tau_1\ (\mathbb{1}_{\mathcal{S}} \otimes \Lambda_{m_k}\,)}{prob\ (m_k)}\ ,$$

where $\mathbb{1}_{\mathcal{S}}$ is the unit operator in $\mathcal{H}_{\mathcal{S}}$ and where $prob\ (m_k) := prob\ (m_k|\tau_1)$ is the probability of obtaining the pointer value $m_k$ given the state $\tau_1$. It may be evaluated:

$$prob\ (m_k) = Tr_{\mathcal{T}}\ \big((\mathbb{1}_{\mathcal{S}} \otimes \Lambda_{m_k}\,)\,\tau_1\big) = \textstyle\sum_i\ (\,|<m^{(i)}\,|\,m_k>|^2\ \ <s_i\,|\,\sigma_0\,|s_i>) =$$

$$= \textstyle\sum_i\ prob\ (m_k\ \big|\ |m^{(i)}>)\ \ \text{x}\ \ prob\,(s_i\,|\,\sigma_0).$$

Note again that the $m_k$ -distribution reflects the $s_i$ -distribution uniquely only if the different wave functions $<m_k\,|\,m^{(i)}>$ do not overlap.

The result for $prob\ (m_k)$ may seem remarkable: no quantum-mechanical interference here! Rather, the probability of finding the value $m_k$ for the pointer variable is the sum over the product of the probabilities $prob\ (m_k\ \big|\ |m^{(i)}>)$ to obtain the pointer value $m_k$, given the state $|\,m^{(i)}>$ that corresponds, in the pre-measurement, to the state $|\,s_i>$, and $prob\ (s_i\ |\sigma_0)$ to obtain the value $s_i$ in the initial state $\sigma_0$ of the system. Indeed this is what one would expect from a pure classical treatment of probabilities, but here it arises from the quantum-mechanical formalism.

So what have we learned concerning the object-system from this read-out of the meter? The answer sits in its density matrix $\sigma_1(\ |m_k)$ after the read-out, representing as it does the state of the system after that action. It is obtained in the usual way by taking the partial trace of the corresponding total density matrix:

$$\sigma_1(\ |m_k) := Tr_{\mathcal{M}}\ \tau_1(\,|m_k) = \frac{1}{prob\ (m_k)}\ Tr_{\mathcal{M}}\,\{(\mathbb{1}_{\mathcal{S}} \otimes \Lambda_{m_k}\,)\ \tau_1\ \ (\mathbb{1}_{\mathcal{S}} \otimes \Lambda_{m_k}\,)\,\} =$$

$$= \frac{1}{prob\ (m_k)}\ <m_k\,|\ \mathcal{U}\ \sigma_0 \otimes \mu_0\ \mathcal{U}^\dagger\,|\ m_k> =$$

$$= \frac{1}{prob\ (m_k)}\ <m_k|\ \mathcal{U}\ |m^{(0)}>\ \sigma_0\ <m^{(0)}\ |\,\mathcal{U}^\dagger\,|\ m_k> = \frac{1}{prob\ (m_k)}\ \ \Omega_k\ \sigma_0\ \Omega_k{}^\dagger\,,$$

where $\Omega_k$ is an operator on $\mathcal{H}_{\mathcal{S}}$ defined by

$$\Omega_k := <m_k\,|\ \mathcal{U}\ |m^{(0)}> = <m_k\,|\ \mathcal{U}\ |m^{(0)}> = \textstyle\sum_i\ <m_k\,|\,m^{(i)}>\ \ |\,s_i><s_i\,|\ =$$

$$= \textstyle\sum_i\ <m_k\,|\,m^{(i)}>\ \ \prod_{s_i}\ \ .$$

The operators $\Omega_k$ are called 'measurement operators' [5]. They play an important role in general theories of measurement [4, 5]. Some of their properties are presented in Appendix 1.

As a final property of the conditional matrix $\sigma_1(\ |m_k)$, note that if we, in describing the system after the pre-measurement, do not care which of the results $m_k$ is obtained in the projective measurement on the meter – we deliberately neglect the outcome of the read-out – then the system must be described by the statistical sum over all possibilities:

$$\sum_k prob\,(m_k) \ \text{x} \ \sigma_1(\ |m_k) = \sum_k \Omega_k\, \sigma_0\, \Omega_k^\dagger =$$

$$= \sum_k < m_k \,|\, \mathcal{U}\, \sigma_0 \otimes \mu_0\, \mathcal{U}^\dagger \,|\, m_k > \ = \ \sigma_1 ,$$

*i.e.* exactly the same unconditional density matrix $\sigma_1$ as the one that describes the system right after the pre-measurement and before the projective measurement of the meter has taken place. For this unconditional density matrix of the system it is thus of no importance whether one performs the projective measurement on the meter or not, provided the result is not recorded. In particular, an un-registered projective meter measurement has no effect on the probability distribution for the eigenvalues $s_i$.

For completeness, I also give the meter's density matrix after the meter has been read-out. It takes the form

$$\mu_1(\ |m_k) = \frac{\Lambda_{m_k}\, \mu_1\ \Lambda_{m_k}}{prob\,(m_k)} \ .$$

In particular, the meter and the object-system become un-entangled after a conditional measurement (but not after an unconditional one).

Now that the machinery is set up, it may be used to answer any legitimate question regarding results of the measurement process. I will use it extensively in the sequel. But first, I will need a further general result.

## *III.D. Consecutive measurements*

For future use, I will consider the case of two successive ancilla measurements. The system remains intact – remember I require non-destructive measurement – but is transformed in steps according to the rules just established.

To start with, let me assume that the second measurement involves a different meter from the first. Looking only at the un-conditional matrices for the system, one has

[5] How to denote a measurement operator varies in the literature, but seems to converge to using $M_k$ instead of what I denote with $\Omega_k$ . Since I use $M$ for entities related to the meter, I have chosen to deviate from this convention.

$$\sigma_0 \xrightarrow[\text{ment}]{\substack{\text{first}\\ \text{measure-}}} \sigma_1 = \sum_{k_1} \Omega_{k_1}^{(1)} \sigma_0 \Omega_{k_1}^{(1)\dagger} \xrightarrow[\text{ment}]{\substack{\text{second}\\ \text{measure-}}} \sigma_2 = \sum_{k_2} \Omega_{k_2}^{(2)} \sigma_1 \Omega_{k_2}^{(2)\dagger} =$$

$$= \sum_{k_1,k_2} \Omega_{k_2}^{(2)} \Omega_{k_1}^{(1)} \sigma_0 \Omega_{k_1}^{(1)\dagger} \Omega_{k_2}^{(2)\dagger} = \sum_{k_1,k_2} (\Omega_{k_2}^{(2)} \Omega_{k_1}^{(1)}) \sigma_0 ( \Omega_{k_2}^{(2)} \Omega_{k_1}^{(1)})^\dagger \ ,$$

*i.e.*, the measurement operators for consecutive ancilla measurement compose naturally. In particular, the generalization to any number of consecutive measurements is obvious.

If the second measurement is identical – has an identical meter to the first and is involved in an identical unitary pre-measurement $\mathcal{U}$ as the first – the expression simplifies to

$$\sigma_2 = \sum_{k,k'} (\Omega_{k'}^{(1)} \Omega_{k}^{(1)}) \sigma_0 ( \Omega_{k'}^{(1)} \Omega_{k}^{(1)})^\dagger = \sum_{i,j} \Pi_{s_i} \sigma_0 \Pi_{s_j} (<m^{(j)}|m^{(i)}>)^2 \ ,$$

as is seen by a short calculation. This expression, too, is easily generalized to any number of identical measurements.

## *IV. Some application of the ancilla scheme*

In this section, I give some examples of applications ('protocols') based on the indirect, or ancilla, measurement scheme. These examples will be used several times later on in the article for more concrete illustration of particular aspects of, *e.g.*, weak measurements and weak values.

### *IV.A. The von Neumann protocol* [6]

Soon after quantum mechanics was invented, von Neumann [1] gave a formulation of the measurement process which is, essentially, a specific application of the ancilla scheme. His protocol gives a more detailed specification of the meter $\mathcal{M}$ and uses a particular form for the interaction Hamiltonian $H_{int}$ between the object-system and the meter and, consequently, for the unitary pre-measurement operator $\mathcal{U}$. I will now describe that measurement protocol, for simplicity for the case where the initial state is a pure state.

The meter is supposed to have a continuous[7] pointer variable $Q$ replacing $M$, and with pointer states $|q>$ replacing $| m_k >$. Almost realistically, $q$ is to be thought of as the pointer position of a graded gauge. One may then introduce the wave functions $\varphi_0(q)$ and $\varphi_i(q)$ from the expansions

$$|m^{(0)}> \ = \ \int dq \ |q> \ <q \ |m^{(0)}> \ =: \ \int dq \ |q> \varphi_0(q) \ ,$$

[6] Do not confuse the von Neumann measurement *protocol* presented here, which is a particular realization of the present ancilla measurement scheme, with the von Neumann *scheme,* which is a term sometimes used as alternative name to the projective measurement scheme described in section II. *C.f.* footnote 1.

[7] I take the usual liberty of the physicist in translating, without further ado, a discrete treatment in the finite-dimensional Hilbert space $\mathcal{H}_{\mathcal{M}}$ into a continuous one.

$$|m^{(i)}> \; = \; \int dq \; |q> \; <q \; |m^{(i)}> \; =: \; \int dq \; |q> \varphi_i(q) \; .$$

I shall assume that the initial pointer state $\varphi_0(q)$ is centered around $q = 0$, *i.e.*, that the expectation value $<Q>_0$ in the initial state vanishes. A particular choice for $\varphi_0(q)$ could be

$$\varphi_0(q) = \{ \frac{1}{\sqrt{2\pi\Delta^2}} \exp(-q^2/2\Delta^2) \}^{½} \; ,$$

so chosen that the probability density $|\varphi_0(q)|^2$ is a Gaussian with width $\Delta$. As will be clear in a moment, the other wave functions, $\varphi_i(q)$, can be expressed in terms of $\varphi_0(q)$.

The next step in this protocol is to specify the interaction Hamiltonian. The choice von Neumann made was

$$H_{int} = \gamma \, S \otimes P.$$

Here, $\gamma$ is a coupling constant and $P$ is the meter variable conjugate to the pointer variable $Q$, *i.e.*, $P$ is the meter momentum, obeying the commutation relation $[\, Q, P \,] = i$ with $Q$ (in units where $\hbar = 1$). Moreover, $S$ is the observable one wants to measure for the object-system; it is as previously assumed to act in a finite-dimensional Hilbert space

The measurement is supposed to occur during a short time interval $\delta t_{\mathcal{U}}$, so that the time integral $\int dt \, H_{int}$ entering in the time-evolution operator $\mathcal{U}$ may be substituted by $\gamma \, S \otimes P \, \delta t_{\mathcal{U}}$; possible effects from switching the interaction on and off is assumed to be taken into account by suitable smoothing of the time-dependence of the interaction[8]. The combination $\gamma \, \delta t_{\mathcal{U}} =: g$ is then a kind of effective coupling constant.

All this, and the simplifying assumption that the intrinsic Hamiltonians $H_{\mathcal{S}}$ and $H_{\mathcal{M}}$ both vanish, imply that the unitary operator $\mathcal{U}$ responsible for the pre-measurement becomes

$$\mathcal{U} = \; exp\,(-\,i \int dt \, H_{\mathcal{T}}) \; = \; exp\,(-\,i \int dt \, H_{int}) \; = \; exp\,(-\,i\,g\,S \otimes P\,)$$

The reason for this very choice of $H_{int}$ hangs on the fact that an operator $exp\,(\,i\,\lambda\,P\,)$, with $\lambda$ a real number, acts as a translation operator on a wave function $\varphi(q)$ in the $q$-basis:

$$exp\,(\,i\,\lambda\,P\,)\,|\,\varphi> \; = \; exp\,(\,i\,\lambda\,P\,) \int dq \;\; |q> \; \varphi(q) \;\; = \int dq \;\; |q> \; \varphi(q+\lambda\,) \; .$$

[8] Another way of doing this is to assume that the coupling 'constant' $\gamma$ is a function $\gamma(t)$ of time $t$ which vanishes for all $t$-values except those near $t = 0$, and such that $\int dt \, \gamma\,(t) \; = \; g$ with $g$ the 'effective' coupling constant (*i.e.*, $\gamma(t)$ would be essentially $\gamma$ times the Dirac $\delta$-function $\delta\,(t)$ ).

Consequently,

$$|s_i > \otimes\ |m^{(0)}> \ \overset{\mathcal{U}}{\rightarrow}\ |s_i > \otimes\ |m^{(i)}> \ = \mathcal{U}\,(\,|s_i> \otimes\ |m^{(0)}>\,) =$$

$$= \ exp\,(-\,i\,g\,S \otimes P\,)\,(\,|s_i > \otimes\ |m^{(0)}>) \ = \ |s_i > \otimes\ (\ exp\,(-\,i\,g\,s_\mathrm{i}\,P\,)\ |m^{(0)}>\,) =$$

$$= \ |s_i > \otimes\ (\,exp\,(-\,i\,g\,s_i\,P\,) \int dq\ |q > \varphi_0(q)\,) \ = \ |s_i > \otimes \ \int dq\ |q > \varphi_0(q - g\,s_i)\,.$$

In other words,

$$|m^{(i)}> \ = \ \int dq\ |q> \ <q\ \ |m^{(i)}> \ = \int dq\ |q > \varphi_i(q) = \int dq\ |q > \varphi_0(q - g\,s_i)\,,$$

so that, finally,

$$\varphi_i(q) = \varphi_0(q - g\,s_i).$$

This means that the initial pointer state of the meter, assumed to be centered around $q = 0$, is transformed by the pre-measurement to a superposition of states $\varphi_i(q)$, which are translations of the initial state by the amount $g\,s_i$. In a very concrete and manifest way this illustrates that the von Neumann protocol describes how the pre-measurement shifts the pointer from $q = 0$ to one of the values $q = g\,s_i$ , which can then be obtained in the read-out. It indeed constitutes a measurement of the system variable $S$!

All relevant quantities related to the measurement may now be evaluated in terms of the shifted states $\varphi_0(q - g\,s_i)$. Let me, *e.g.*, consider the meter density matrix after the pre-measurement:

$$\mu_1 = \ \ \Sigma_i\ |m^{(i)}> \ < s_i\,|\,\sigma_0\,|s_i> \,< m^{(i)}\,| =$$

$$= \Sigma_i\ < s_i\,|\,\sigma_0\,|s_i> \int dq\,dq'\ \ |q> \ \ \varphi_0(q - g\,s_i)\ \ \varphi_0(q' - g\,s_i)^* \ \ <q'|\,.$$

It can be used, *e.g.*, to calculate the mean value $<h(Q)>_1$ of any function $h(Q)$ of the pointer variable $Q$ after the pre-measurement (indicated by the subscript 1 ):

$$< h(Q)>_1 \ := \ Tr\,[h(Q)\,\mu_1) = \Sigma_i\ < s_i\,|\sigma_0\,|s_i> \ \int dq \ \ h(q)\,|\,\varphi_0\,(q - g\,s_i)\,|^2.$$

In particular (as usual, a subscript 0 indicates entities in the initial state),

$$<Q>_1 \ = \ \Sigma_i\ < s_i\,|\sigma_0\,|s_i> \ \int dq \ \ q\,|\,\varphi_0(q - g\,s_i)\,|^2$$

$$= g < S >_0 \ + <Q>_0 \ \ = \ g < S >_0\ ,$$

since $<Q>_0 = 0$ by assumption, and

$$<Q^2>_1 \ = \ <Q^2>_0 \ \ + g^{\,2}< S^2>_0 \ \ ,$$

so that the variance (‘spread’) of the $Q$-distribution after the pre-measurement is augmented compared to the initial state by the variance of the $S$-distribution:

$$\Delta Q_1{}^2 := \langle Q^2\rangle_1 - (\langle Q\rangle_1)^2 = \Delta Q_0{}^2 + g^2 \Delta S_0{}^2 ,$$

a very reasonable result.

## *IV. B. The double qubit*

By the ingenuity of the experimentalists it is possible to produce single photons with well-defined polarization, to entangle two photons prepared in that way, and to measure the polarization of either or both [16]. This type of experiment in which one qubit measures another qubit is aptly amenable to analysis in the indirect measurement scheme of this section [17].

In such a photon experiment one describes the photon polarization states in a 2-dimensional Hilbert space in terms of basis vector $|H\rangle$ and $|V\rangle$, corresponding respectively to horizontal and vertical polarizations. Essentially only for ease of notation, I shall instead use a so called “computational basis” in which $|H\rangle$ is substituted by $|0\rangle$ and $|V\rangle$ by $|1\rangle$ ([18,19]). I shall attach a subscript $\mathbb{S}$ for entities belonging to the system (or ‘signal’) photon, a subscript $\mathcal{M}$ for those belonging to the meter (or ‘probe’) photon (except towards the end of the subsection where only entities referring to the signal photon Hilbert space occur, so no subscript will be needed).

The translation of the general scheme into this particular realization then takes the following form.

For the initial states one chooses

$$|s\rangle = \alpha\,|0\rangle_{\mathbb{S}} + \beta\,|1\rangle_{\mathbb{S}}, \quad \alpha, \beta \text{ complex numbers, } |\alpha|^2 + |\beta|^2 = 1 ,$$

$$|m^{(0)}\rangle = \cos\vartheta/2\ |0\rangle_{\mathcal{M}} + \sin\vartheta/2\ |1\rangle_{\mathcal{M}} .$$

The observable to be measured is $S_1{}^{\mathbb{S}} = |0\rangle_{\mathbb{S}}\,{}_{\mathbb{S}}\langle 0| - |1\rangle_{\mathbb{S}}\,{}_{\mathbb{S}}\langle 1|$, which in the photon case is (one of ) the so called Stokes’ parameter(-s); here, it may be identified with the Pauli matrix $\sigma_z$ (no relation whatsoever to my notation for the system density matrix!). Its eigenstates are $|0\rangle_{\mathbb{S}}$ and $|1\rangle_{\mathbb{S}}$ with eigenvalues +1 and – 1, respectively.[9]

For the pre-measurement $\mathcal{U}$, the protocol says

[9] Note that my use here of the computational basis deviates from my notation elsewhere in the article. In a fully consistent notation, $|0\rangle_{\mathbb{S}}$ should be replaced by $|s_i{=}{+}1\rangle$ and $|1\rangle_{\mathbb{S}}$ by $|s_i{=}{-}1\rangle$. I hope the reader can bear these inconsistencies.

$$|0>_{\mathbb{S}} \otimes |k>_{\mathcal{M}} \overset{\mathcal{U}}{\to} |0>_{\mathbb{S}} \otimes |k>_{\mathcal{M}} ,$$

$$|1>_{\mathbb{S}} \otimes |k>_{\mathcal{M}} \overset{\mathcal{U}}{\to} |1>_{\mathbb{S}} \otimes |1-k>_{\mathcal{M}} , \; k = 0,1$$

.

(In computer language jargon, this is called an ‘CNOT’ gate ). For an arbitrary (but pure) initial state the pre-measurement then implies

$$|s> \otimes |m^{(0)}> = ( \alpha |0>_{\mathbb{S}} + \beta |1>_{\mathbb{S}} ) \otimes$$

$$\otimes (\cos \vartheta/2 \; |0>_{\mathcal{M}} + \sin \vartheta/2 \; |1>_{\mathcal{M}} ) \overset{\mathcal{U}}{\to}$$

$$\overset{\mathcal{U}}{\to} \alpha |0>_{\mathbb{S}} \otimes (\cos \vartheta/2 \; |0>_{\mathcal{M}} + \sin \vartheta/2 \; |1>_{\mathcal{M}} ) +$$

$$+ \beta |1>_{\mathbb{S}} \otimes (\sin \vartheta/2 \; |0>_{\mathcal{M}} + \cos \vartheta/2 \; |1>_{\mathcal{M}} ),$$

From here, one identifies the two possible meter states after the pre-measurement:

$$|m^{(s_i = +1)}> := \cos \vartheta/2 \; |0>_{\mathcal{M}} + \sin \vartheta/2 \; |1>_{\mathcal{M}} ,$$

$$|m^{(s_i = -1)}> := \sin \vartheta/2 \; |0>_{\mathcal{M}} + \cos \vartheta/2 \; |1>_{\mathcal{M}} ,$$

which in their turn can be used to obtain, *e.g.*, the total system density matrix $\tau_1$ .

## *IV.C. Amplification by post-selection*

In the ideal measurement scheme, restricting the events under consideration to one particular final state$| f >$ for the system by ‘post-selection’, led to the ABL rule as spelled out in section II.C. Since the ancilla scheme does not measure the system observable *S* directly, it does not provide any direct way to obtain the ABL rule. But in the limit of a strong measurement – which, as shown in section III.B, can be achieved if there is no overlap between the different meter states $|m^{(i)} >$ – one retrieves all the results of the projective scheme, including the ABL rule.

In the ancilla scheme, there are more degrees of freedom than those for the system, *viz.*, those for the meter. This may be used to extract new information. In particular, a judicious choice of a post-selected state $| f >$ for the system with respect to the pre-selected one, $\sigma_0$ (or $| s > < s |$ if it is a pure state), provides a technique for amplification that has been exploited in exquisite ways by experimentalists. I devote this section to the theoretical background for that technique, while section VII.B describes the experiments. Similar considerations may be found in [20-22]

To exploit the added degrees of freedom of the meter, the trick in the ancilla scheme is, firstly, to make *no* read-out on the meter: all density matrices should be the unconditional ones. In particular, the joint system density matrix after pre-measurement becomes

$$\tau_1 = \Sigma_{i,j} \; (| \, m^{(i)} > \Pi_{s_i} \, \sigma_0 \; \Pi_{s_j} < m^{(j)} | \;) \; .$$

Next, post-select $| \, f >$ for the system, turning the joint system density matrix $\tau_1$ into

$$\tau_f := \frac{(\, \Pi_f \otimes \mathbb{1}_{\mathcal{S}}) \;\, \tau_1 \;\, (\, \Pi_f \otimes \mathbb{1}_{\mathcal{S}})}{prob \; (f \, | \, \tau_1)} \quad ,$$

where

$$prob \; (f \, | \, \tau_1) \; = \; Tr \, \big( (\, \Pi_f \otimes \mathbb{1}_{\mathcal{S}}) \;\, \tau_1 \, \big).$$

Finally, focus on the meter density matrix $\mu_f := Tr_{\mathcal{S}} \; \tau_f$ after the post-selection, and use it to investigate suitable observables for the meter in that state, trying to get the amplifying effect from choosing the denominator $prob \; (f \, | \, \tau_1)$ as small as possible.

I shall not treat the most general case, but start by concentrating on the special case where the system is a qubit, described by a pure state

$$|s > \; = \alpha \, |0 > \; + \; \beta \, |1 > , \quad \alpha, \beta \text{ complex numbers, } |\alpha|^2 \; + |\beta|^2 \; = 1.$$

The system observable is chosen to be

$$S \; = \; |0><0| \; - \; |1><1| \, ,$$

*i.e.*, one of the Stokes parameters in case the qubit is a photon (*c.f.* section IV.B).

As for the meter, it is taken to be a von Neumann one (see section IV.A), so that

$$\mathcal{U} = exp \; (- \, i \, g \, S \otimes P \; )$$

with $P$ the meter operator conjugate to the pointer variable $Q$. Consequently,

$$|s > \otimes \; | \, m^{(0)} > \; = \; |s > \otimes \int dq \; |q > \varphi_0(q) \; \overset{\mathcal{U}}{\rightarrow}$$

$$\overset{\mathcal{U}}{\rightarrow} \; \alpha \, |0 > \int dq \; |q > \varphi_0(q - g) \; + \; \beta \, |1 > \int dq \; |q > \varphi_0(q + g) \; .$$

In fact, this is a good model for a Stern-Gerlach set-up, with $g$ a measure of the (inhomogeneous) magnet field – in fact, $g$ is proportional to $\mu \frac{\partial B_z}{\partial x}$ where $\mu$ is the magnetic moment of the electron (no connection whatsoever to my notation for the meter density matrix!) and $B_z$ the magnetic field causing the spin components in the *z*-direction to be separated – and $Q$ the length coordinate along this *z*-direction[10].

Upon post-selection of a system state $| f >$ (in the Stern-Gerlach set-up, this should be the eigenstate of a spin component other than the ones in the z-direction), the matrix element of the ensuing meter final density matrix $\mu_f$, calculated as described above, may be written

$$< q \,|\, \mu_f \,|\, q' > \;=\; \frac{1}{prob\;(f\,|\,\tau_1)}\; \tilde{\varphi}_f(q)\; \tilde{\varphi}_f(q')^* \,.$$

Here, with the abbreviations $\alpha_f := \alpha < f \,|\, 0 >$ and $\beta_f := \beta < f \,|\, 1 >$,

$$\tilde{\varphi}_f(q) := \alpha_f\; \varphi_0(q-g) \;+\; \beta_f\; \varphi_0(q+g)$$

is the (non-normalized) wave function for the meter after the post-selection and

$$prob\;(f\,|\,\tau_1) \;=\; \int dq\; |\tilde{\varphi}_f(q)|^2 \;=$$

$$=\; |\alpha_f|^2 \;+ |\beta_f|^2 \;+\; 2\; \mathcal{R}e \left( \int dq\;\; \alpha_f\, \varphi_0(q-g)\, \beta_f^*\; \varphi_0(q+g)^* \right).$$

In fact, the expression for the wave function may be derived directly from considerations of the state vectors instead of using density matrix formalism.

Of particular interest shall be the mean value $_f\!< Q >$ of the pointer variable after the post-selection. A short calculation gives

$$_f\!< Q > \;:=\; Tr\,(Q\; \mu_f\,) \;=\; \frac{g\;(\;|\alpha_f|^2 - |\beta_f|^2\;)}{|\alpha_f|^2 + |\beta_f|^2 + 2\; r\; \mathcal{R}e\; \alpha_f\; \beta_f^*}$$

where

$$r := \int dq\;\; \varphi_0(q-g)\; \varphi_0(q+g)^*,$$

the overlap between the two wave functions, is assumed to be real (which it is , *e.g.*, for $\varphi_0(q)$ an even function of its argument $q$). [This simple form for $Tr\,(Q\;\mu_f\,)$ also requires cross products like $\int dq\; q\; \varphi_0(q-g)\; \varphi_0(q+g)^*$ to vanish, which would occur, *e.g.*, for $\varphi_0(q)$ an even function of its argument $q$ and real.]

The expression for $_f\!< Q >$ is to be compared to the mean value $< Q >_1$ of the pointer variable without any post-selection,

$$< Q >_1 \;:=\; Tr\;(\;Q\;\mu_1\;) = g\;(\;|\,\alpha\,|^2 \;-\; |\,\beta\,|^2\;)\,.$$

[10] An even better model of the Stern-Gerlach set-up would be to interchange the roles of the variables *P* and *Q*. However, the mathematics is the same, so I stick to the way I have introduced the von Neumann protocol.

One realizes immediately the possibility of reaching much larger values for ${}_f\!< Q >$ than for $< Q >_1$ due to the denominator $prob\ (f\,|\,\tau_1)$ ) in the expression for ${}_f\!< Q >$. Upon choosing a small value for $prob\ (f\,|\,\tau_1)$, one could expect a large amplification effect, ${}_f\!< Q > / < Q >_1$ >> 1. In other words, accommodating the post-selected state $|\,f >$ so that only rare events are considered – low $prob\ (f\,|\,\tau_1)$ – one might hope to achieve a large amplification.

A more concrete result is obtained by studying ${}_f\!< Q >$ as a function of $< f\,|\,s >$. A straight-forward calculation, assuming $\alpha_f \neq 0$, shows that

$$|\,{}_f\!< Q >| \leq g\,/\,\sqrt{1-r^2}\ ,$$

with equality for

$$< f\,|\,s > \ = \ \alpha_f\ \frac{1}{r}\ (\pm\ \sqrt{1-r^2}\ -\ (1-r)\,).$$

For that value of $< f\,|\,s >$ one has

$$prob\ (f\,|\,\tau_1)\ =\ 2\,\frac{|\alpha_f|^2\ (1-r^2)}{1+\sqrt{1-r^2}}\ .$$

As one sees, the amplification effect becomes larger, the nearer $r$ is to $\pm$ 1, but with a corresponding decrease of the probability to find the state $|\,f >$. This leads me to consider, in the next section, so called weak measurements, in which one strives for r → 1.

But before that, let me quote two other results.

Firstly, it is also of some interest to calculate the variance ("spread"), $\Delta_f Q^{\,2}$, of the pointer distribution after the post-selection. For a Gaussian initial meter function, one finds

$$\Delta_f Q^{\,2}\ =\ \Delta Q_0^{\,2}\ +\ g^2\left(\frac{|\alpha_f|^2+|\beta_f|^2}{prob\ (f\,|\,\tau_1)} - \frac{(\,|\alpha_f|^2-|\beta_f|^2\,)^2}{(prob\ (f\,|\,\tau_1))^2}\right).$$

For ${}_f\!< Q >$ at the maximal value one has

$$\Delta_f Q^{\,2}\ =\ \Delta Q_0^{\,2}\ .$$

At maximal amplification, the spread of the final pointer distribution is thus the same as the corresponding initial one, independent of the strength of the measurement. In fact, there are even values of $< f\,|\,s >$ for which the final pointer spread is smaller than the initial one.

Secondly, I will just mention the results of the amplification considerations for the double qubit of section IV.B. Very similar arguments as those for the von Neumann protocol above, and without extra assumptions, lead to

$$prob\ (f\,|\,\tau_1)\ = |\alpha_f|^2 + |\beta_f|^2\ +\ 2\ r_{dq}\ \mathcal{R}e\,(\alpha_f\ \beta_f{}^*)\ ,$$

with $r_{dq} = \sin\vartheta$ , the angle $\vartheta$ giving the initial composition of the meter qubit as given in section IV.B. Consequently, the mean value ${}_f\langle S_1^{\mathcal{M}}\rangle$ of the Stokes' parameter

$$S_1^{\mathcal{M}} := |0\rangle_{\mathcal{M}\ \mathcal{M}}\langle 0| - |1\rangle_{\mathcal{M}\ \mathcal{M}}\langle 1|$$

for the meter qubit, after the post-selection as above on $|f\rangle$ for the system, becomes

$${}_f\langle S_1^{\mathcal{M}}\rangle := Tr\,(S_1^{\mathcal{M}}\mu_f) = \cos\vartheta\ \left(|\alpha_f|^2 - |\beta_f|^2\right) / \,prob\,(f\,|\,\tau_1)\ .$$

The similarity of this expression to the one for ${}_f\langle Q\rangle$ above, allows one to draw the conclusion that

$$|\,{}_f\langle S_1^{\mathcal{M}}\rangle\,| \leq \cos\vartheta \,/\, \sqrt{1-\sin\vartheta^{\,2}} = 1.$$

Consequently, and maybe not too astonishing: from the point of view of amplification in the double qubit case, there is no real gain using post-selection in the sense that one will never reach outside the conventional interval.

## *V. Weak measurements – with or without post-selection*

A quantum mechanical measurement, described either by the ideal measurement scheme of section II or the indirect one of section III, in general implies that the object-system under study will suffer large changes ('disturbances') in its state, even if the measurement is considered non-destructive; indeed, non-destructive only means that the system is not annihilated in the measurement. In the formalism I have described, these disturbances appear in the change from the initial density matrix $\sigma_0$ to a usually quite different matrix $\sigma_1$ after the measurement.

A weak measurement is a measurement that disturbs the state of the object of interest as little as possible. As we will see, a weak measurement is also such that the measurement results are less clear than in a 'strong' or 'sharp' measurement. For example, there will be difficulties in distinguishing one eigenvalue of the observable under study from another. This feature, however, can and has to be compensated for by 'increasing the statistics', *i.e.*, repeating the experiment many more times.

But is it not kind of silly to deliberately refrain from doing a measurement as 'sharply' as possible? The answer is that new phenomena occur for weak measurement, phenomena that can only be studied by weakening the interaction responsible for the measurement as much as possible. As has been advocated, in particular by Aharonov and Vaidman and their collaborators, the idea of using weak measurement is particularly fruitful when combined with 'post-selection'. Let me first give a short overview of these ideas.

## *V.A. Preliminaries on weak values from indirect measurement with post-selection*

Consider a set-up like the one I described in section III, with the system under study coupled to a measurement device, the meter. In this preliminary presentation, let me focus on the more concrete expression for $\mathcal{U}$ from the von Neumann protocol of section IV.A. For simplicity, let the system initial state be a pure one, $|s>$. What I called the pre-measurement $\mathcal{U}$ amounts to the transition

$$|s>\otimes\ |m^{(0)}> \overset{\mathcal{U}}{\rightarrow}\ \mathcal{U}(\ |s>\otimes|m^{(0)}>\ )\ = exp\,(-\,i\,g\,S\otimes P\,)\,(\ |s>\otimes\ |m^{(0)}>\,)\,.$$

A weak measurement may be characterized by the fact that $g$ is small[11], so that a Taylor series expansion of the exponential, $exp\,(-\,i\,g\,S\otimes P\,)\ \approx\ (1-i\,g\,S\otimes P\,)$, should be valid. By post-selecting on a state $|f>$ for the system, and assuming $<f\,|s>\ \neq 0$, the resulting final state $|m^{(f)}>$ for the meter may be written

$$|m^{(f)}>\ :=\ <f\,|\ exp\,(-\,i\,g\,S\otimes\,P\,)\,(|s>\otimes|m^{(0)}>\,)\approx$$

$$\approx\ <f\,|s>\Big(\,1-i\,g\ \frac{<f\,|\,S\,|s>}{<f\,|\,s>}\ P\,\Big)\,|m^{(0)}>\ \approx\ <f\,|s>\ \ exp\,(-\,i\,g\,S_{\rm w}\,P\,)\,|m^{(0)}>\ ,$$

where the so called ‘weak value’ $S_{\rm w}$ of $S$ is defined by

$$\text{weak value of } S :=\ \ \frac{<f\,|\,S\,|s>}{<f\,|\,s>}\ \ =:\ S_{\rm w}\,.$$

Moreover, and for simplicity, $S_{\rm w}$ has been assumed real; this will be relaxed in due course.

It follows from the treatment of the von Neumann protocol in section III.B that the result for $|m^{(f)}>$ means that the initial meter wave function $<q|m^{(0)}>\ = \varphi_0(q)$, centered around $q = 0$, is transformed into a final wave function

$$\varphi_f(q)\ :=\ <q\,|m^{(f)}>\ = \varphi_0(q-g\,S_{\rm w}\,)\,,$$

i.e., the initial one shifted by $g\,S_{\rm w}$ and not by any of the eigenvalues $g\,s_i$, as was the case in the original von Neumann protocol.

This procedure thus puts focus on a new entity for the object-system $\mathbb{S}$, *viz*., the weak value $S_{\rm w}$ of an observable. Note that $S_{\rm w}$ depends both on the initial system state $|s>$ and on the final one, $|f>$. Note also that the way it enters the game is via an ancilla measurement. There is no room for it in a projective measurement scheme, simply because there is no 'weak measurement' to be defined in that scheme.

---

[11] Aharonov *et al* have a slightly different approach, essentially amounting to – in my conventions – characterizing a weak measurement as one in which the spread $\Delta P_0$ in the meter momentum observable $P$ tends to zero, equivalent to a spread $\Delta Q_0$ in the pointer variable $Q$ that tends to infinity. The two approaches are equivalent whenever both apply; in fact, the relevant small parameter is rather $g/\Delta Q_0$. *C.f.* also footnote 12 below.

The weak value has been the focus of much almost philosophical debate; see, *e.g.*, [7] and references therein. It has also been used in several ingenious ways by experimentalists – see section VII below – to study entities not thought measureable at all, like the trajectories in a double slit experiment or the wave function of an object; it has also been used for amplification purposes. Before I comment a little on the meaning of a weak value, and describe some of these experimental items, I must elaborate on the rather crude derivation just given and look a little bit more in detail on how weak measurements in general are to be described.

## *V.B. Weak measurement in the ancilla scheme.*

As I already emphasized, weak measurements must be treated in the ancilla scheme.

From section III, recall that the total system density matrix $\tau_1$ after the pre-measurement $\mathcal{U}$ is $\tau_1 = \mathcal{U}\ \tau_0\ \mathcal{U}^\dagger$ . Here, $\tau_0 = \sigma_0 \otimes \mu_0$ is the direct product of the system initial density matrix $\sigma_0$ and the meter initial density matrix $\mu_0 = |m^{(0)}> < m^{(0)}|$ . Now, assuming that in $\mathcal{U} =$ $= exp\,(-i \int dt\, H_{\mathcal{T}})$ we may write (*c.f.* the treatment of the von Neumann protocol in section IV.A; also, recall that I assume the intrinsic Hamiltonians $H_{\mathbb{S}}$ and $H_{\mathcal{M}}$ to vanish)

$$exp\,(-i \int dt\, H_{\mathcal{T}}) = exp\,(-i \int dt\, H_{int}) = exp\,(-i\, g\, S \otimes N),$$

with $g$ an (effective) coupling constant, $S$ the operator/observable under study for the object-system, and $N$ a Hermitian operator , a (conjugate) 'pointer variable' in the Hilbert space $\mathcal{H}_{\mathcal{M}}$ of the meter. Besides being motivated by the choices made in the von Neumann protocol, one may also argue that the interaction Hamiltonian, representing, as it should a non-destructive measurement of $S$, should commute with $S$. The simplest choice is then to have it proportional to that operator. The meter operator $N$ will be specified later; it generalizes the $P$ operator of the von Neumann protocol. For ease of writing, I also temporarily introduce the abbreviation $\mathcal{X} :=$ $= g\, S \otimes N$. Then[12]

$$\mathcal{U} = exp\,(-i \int dt\, H_{\mathcal{T}}) = exp\,(-i\,\mathcal{X}) = 1 - i\,\mathcal{X} - \tfrac{1}{2}\mathcal{X}^2 + O(\mathcal{X}^3)\ ,$$

so that (an approximate sign, $\approx$ , means equality disregarding terms of order $\mathcal{X}^3$ and higher)

[12] This is of course sloppy mathematics (but no worse than what is done by most physicists on many other occasions): one would also need to have the operators $S$ and $N$ in some sense small. I shall rely on the naïve physicist's intuition, hoping that a more strict mathematical treatment will be able to justify the approximations made. I note, however, that given the legitimacy of the expansion and the following deductions, what shall be required to be small are products like $g <S>_0\ (<N^2>_0)^{½}$ (notation not defined follows below).

$$\tau_1 = \mathcal{U}\ \tau_0\ \mathcal{U}^\dagger \approx (1 - i\mathcal{X} - \tfrac{1}{2}\mathcal{X}^2)\ \tau_0\ (1 + i\mathcal{X} - \tfrac{1}{2}\mathcal{X}^2) \approx$$

$$\approx \tau_0 + i\,[\tau_0, \mathcal{X}] - \tfrac{1}{2}(\mathcal{X}^2\ \tau_0 + \tau_0\ \mathcal{X}^2 - 2\,\mathcal{X}\,\tau_0\,\mathcal{X}) = \tau_0 + i\,[\tau_0, \mathcal{X}] - \tfrac{1}{2}[\,[\tau_0, \mathcal{X}]\,, \mathcal{X}\,]\ ,$$

with brackets [ … ] symbolizing a commutator; the last term is then a double commutator.

Using the simplifying assumption $< N >_0 := Tr_{\mathcal{M}}(\mu_0 N) = 0$ (but $<N^2>_0 := Tr_{\mathcal{M}}(\mu_0 N^2) \neq 0$ ) the system's unconditional density matrix reads

$$\sigma_1 = Tr_{\mathcal{M}}\,\tau_1 \approx \sigma_0 - \tfrac{1}{2}\ g^2 <N^2>_0\ \ [\,[\sigma_0, S\,]\,, S\,]\ .$$

This becomes more illuminating if one exhibits its matrix elements in the $|s_i>$-basis:

$$< s_i\ |\sigma_1\ |s_j> \approx\ <s_i\ |\sigma_0\ |s_j> \left(1 - \tfrac{1}{2}\ g^2 <N^2>_0\ \ (s_i - s_j)^2\right),$$

an expression which indeed shows characteristics of a weak measurement: the correction to the initial density matrix is even of second order in the strength $g$ of the measurement.

Next, consider the overlap $< m^{(i)}|m^{(j)}>$ of the final meter state in this approximation. A short calculation – or a direct comparison of the expression for $< s_i\ |\sigma_1|s_j>$ just derived with the general expression in section III.B – shows that

$$< m^{(i)}|m^{(j)}>\ \approx\ 1 - \tfrac{1}{2}\ g^2 <N^2>_0\ \ (s_i - s_j)^2,$$

meaning that for a weak measurement there is (almost) total overlap between the different meter states, *i.e.*, they do not any longer differentiate (effectively) between the different $s_i$ -values. This is again a special characteristic of weak measurements.

The final meter state, calculated in the same approximation, is

$$\mu_1 = Tr_{\mathbb{S}}\,\tau_1 \approx \mu_0 + i\,g < S >_0\ \ [\mu_0, N\,] - \tfrac{1}{2}\ g^2 <S^2>_0\ \ [\,[\mu_0, N\,]\,, N\,]\ .$$

Let me immediately note one consequence of this result. I want to find the after-measurement mean value, $<M>_1$ , of the meter variable $M$ conjugate to $N$ – meaning that the commutator $[M\ , N] = i$ ; of course, this is copied on the von Neumann protocol case, where one identifies $N$ with $P$ and $M$ with $Q$. Assuming the corresponding mean value $<M>_0$ in the initial state to vanish, it reads

$$<M>_1\ =\ g < S >_0\ + O(g^2)\ .$$

This relation between the mean values of the pointer variable $M$ and the system variable $S$ is a further characteristic of weak measurements.

## *V. C. Weak measurement followed by post-selection.*

The next step in the protocol is to subject the system after the weak measurement, when it is described by a total density matrix

$$\tau_1 \approx \tau_0 + i\,[\tau_0, g\,S \otimes N] - ½\,[\,[\tau_0, g\,S \otimes N]\,, g\,S \otimes N\,]\,,$$

to a post-selection on the state $|f>$ for the system. The procedure is graphically illustrated in figure 4. Since the object-system and the meter are still entangled before this action – it is assumed that no read-out of the meter has taken place – the meter state $\mu_f$ will also be influenced by the post-selection. It takes the form

$$\mu_f := \frac{1}{prob\,(f\,|\,\tau_1)}\; Tr_{\mathbb{S}}\,(\,\Pi_f \otimes \mathbb{1}_{\mathcal{M}}\;\tau_1\,)$$

Here $\Pi_f = |f><f|$ is the usual projector, so that the numerator reads

$$Tr_{\mathbb{S}}\,(\,|f><f| \otimes \mathbb{1}_{\mathcal{M}}\;\tau_1\,) \approx <f|\,\sigma_0\,|f>\;\mu_0 + i\,g\,\{\,<f|\,\sigma_0\,S\,|f>\,\mu_0\,N - <f|\,S\,\sigma_0\,|f>\;N\,\mu_0\,\} -$$

$$- ½\;g^2\,\{\,<f|\,\sigma_0\,S^2|f>\,\mu_0\,N^2 + <f|\,S^2\,\sigma_0\,|f>\;N^2\mu_0 - 2<f|\,S\,\sigma_0\,S\,|f>\;N\,\mu_0\,N\,\}\,.$$

For ease of comparison with what is usually quoted in the literature, I shall in the following only consider the case of a pure initial system state, $\sigma_0 = |s><s|$. Then

$$Tr_{\mathbb{S}}\,(\,|f><f| \otimes \mathbb{1}_{\mathcal{M}}\;\tau_1) \approx |<f|s>|^2\,\mu_0 + 2\,g\,\text{“}\mathcal{I}m\text{”}\,\big(<f|\,S\,|s><s\,|f>\;N\,\mu_0\big) -$$

$$- g^2\,\text{”}\mathcal{R}e\text{”}\,\big(<f|\,S^2|s><s\,|f>\;N^2\mu_0 - <f|\,S\,|s><s\,|S\,|f>\;N\,\mu_0\,N\big).$$

Here, the symbol “$\mathcal{I}m$” (respectively ”$\mathcal{R}e$”) means the anti-Hermitian (respectively Hermitian) part of the bracket that follows. Assuming $<f|s> \neq 0$, this may be written

$$Tr_{\mathbb{S}}\,(|f><f| \otimes \mathbb{1}_{\mathcal{M}}\;\tau_1\,) \approx$$

$$\approx |<f|s>|^2\,\big(\mu_0 + 2\,g\,\text{“}\mathcal{I}m\text{”}\,(S_{\rm w}\,N\,\mu_0) - g^2\,\text{”}\mathcal{R}e\text{”}\,(\,\frac{<f|S^2|s>}{<f|s>}\;N^2\mu_0 - |S_{\rm w}|^2\;N\,\mu_0\,N)\big)\,,$$

where $S_{\rm w}$, as introduced above, is the weak value of the observable $S$.

The expression for the probability $prob\,(f\,|\,\tau_1)$ is most easily obtained from the requirement that $Tr\,\mu_f = 1$. It reads, to second order in the strength $g$ of the interaction, and using $<N>_0 = 0$,

$$prob\,(f\,|\,\tau_1) \approx |<f|s>|^2\,\big(1 - g^2<N^2>_0\;\mathcal{R}e\,(\,\frac{<f|S^2|s>}{<f|s>} - |S_{\rm w}|^2)\big).$$

It pays to stop for a while and digest this result. It says that the probability to obtain a final state $|f>$ after the pre-measurement, and for a system initially in a state $|s>$, is, to leading order in the measurement strength $g$, given by the overlap $|<f|s>|^2$. This is not that astonishing. But it is important for the subsequent applications of the procedure, in which the weak value $S_{\rm w}$ of the observable $S$ for the system is in focus. The case of amplification is particularly important. In the special cases I studied in section IV.C, large amplification requires a small value of $prob\,(f\,|\,\tau_1)$, which for weak measurement means a small value of $< f\,|s >$, in turn implying a large value of $S_w$. I will treat amplification in the weak measurement scheme in more detail in section V.E below.

Continuing on the main track, the final expression for the meter density matrix $\mu_f$ after post-selecting on the state $|f>$ – assuming the initial object state to be pure, $\sigma_0 = |s><s|$, that $< f\,|s > \neq 0$, and correct to second order in the interaction strength $g$ – then reads

$$\mu_f \approx \big(\mu_0 + 2\,g\,“\mathcal{Im}”\,(S_w\,N\,\mu_0\,) - g^2\,”\mathcal{Re}”\,(\,\frac{<f|S^2|s>}{<f|s>}\;N^2\mu_0 - |S_w|^2\;N\,\mu_0\,N)\big)\,/\mathcal{D}.$$

For easier writing, I here introduced the abbreviation

$$\mathcal{D} \;:=\; 1 \;-\; g^2 <N^2>_0\;\mathcal{Re}\,(\frac{<f|S^2|s>}{<f|s>} \;-\; |S_w|^2)\;,$$

where $\mathcal{Re}$ (and subsequently $\mathcal{Im}$) now means the ordinary real (respectively imaginary) value. As is seen, the weak value $S_w$ takes a prominent place here; in fact, to first order in the interactions strength $g$ it is the only reference to the object that remains.

The expression for $\mu_f$ contains all relevant information of the joint meter-system state after the post-selection; the meter and the object-system are of course no longer entangled after the post-selection. It may be used to calculate any relevant quantity one wants. In that way, one may deduce the value of $S_w$ from suitably chosen measurements on the meter. In this sense, $S_w$ is a measurable quantity.

More concretely, let $L$ be any meter observable. Neglecting the $g^2$ term in the nominator, one has for the expectation value ${}_f\!< L >$ of $L$ after the post-selection

$${}_f\!< L > \;:=\; Tr\,(\,L\,\mu_f) \;\approx\; < L >_0 \;+\; 2\,g\,\mathcal{Im}\,(\,S_w < L\,N\,\mu_0 >)\,/\,\mathcal{D}\,.$$

Here, it is instructive to write $L\,N \;=\; ½\,(\,[L\,,N] + \{L\,,N\,\}\,)$, *i.e.*, as a sum of a commutator and an anticommutator, so that

$$2\;\mathcal{Im}\,(S_w < L\,N >_0) = -\,i\;\;< [\,N\,,L\,] >_0\,\mathcal{Re}\,S_w + < \{\,N\,,L\,\} >_0\;\mathcal{Im}\,S_w\;\;,$$

explicitly exhibiting the real and imaginary parts of $S_w$ separately.

Two concrete choices of $L$ may illustrate the procedure. Consider first $L = N$. Then, within the approximations made,

$${}_f\!< N > \approx 2\,g\;\; < N^2 >_0\;\;\mathcal{Im}\;S_w\,/\;\mathcal{D}.$$

The second case is the same as was considered in subsection V.B above, *viz.*, of an operator $M$ that is conjugate to the operator $N$ in the sense that the commutator $[M, N\,] = i$ ; the anticommutator $\{\,M\,,N\,\}$ in general remains unknown. It follows that

$${}_f\!< M > \approx (\;g\;\mathcal{Re}\,S_w + g < \{\,N\,,\,M\} >_0\;\mathcal{Im}\,S_w\;\;)\,/\;\mathcal{D}\,,$$

so from here $\mathcal{R}e\, S_w$ may be recovered as well.

## *V. D. Application 1: Weak measurement in the von Neumann protocol*

Much of the treatment above is tailored on the von Neumann measurement protocol of section IV.A. So it is not astonishing that the results just derived may be directly applied to that case.

I first note that the wave function of the meter after the pre-measurement but before post-selection becomes

$$\varphi_1(q) \approx \varphi_0(q - g <s|S|s>)$$

reflecting the fact that $<Q>_1 = g <S>_0 = <s|S|s>$ (c.f. the expression for $<M>_1$ in section V.B.)

Let us stop for a moment to comprehend this result. As a mean value, $<s|S|s>$ is usually thought of as the sum of its eigenvalues weighted with their respective probabilities. To measure it would, then, require many measurements on the system $\mathbb{S}$ to find these eigenvalues and probabilities. Here, it appears in principle directly from the (weak) measurement on the ancilla through the wave function of the latter. Of course, there are caveats: In any case, it will be a matter of getting the statistics. And is it really easier or more economical to make the necessary (weak) measurement on the ancilla than to determine $<s|S|s>$ in the conventional way?

From the results in section V.C, with the substitions $N \hookrightarrow P$ and $M \hookrightarrow Q$ , one immediately deduces

$$_f\!< Q > \approx \big( g\, \mathcal{R}e\, S_w + < \{ P , Q \} >_0 \mathcal{I}m\, S_w \big) / \mathcal{D}_{vN}$$

and

$$_f\!< P > \approx 2\, g\, < P^2 >_0 \mathcal{I}m\, S_w \;/\; \mathcal{D}_{vN}$$

with

$$\mathcal{D}_{vN} := 1 - g^2 <P^2>_0 \mathcal{R}e\,\Big( \frac{<f|S^2|s>}{<f|s>} - |S_w|^2 \Big),$$

Note in particular that $_f\!< Q >$ contains a term $\propto \mathcal{I}m\, S_w$ : in general, $< \{ P , Q \} >_0 \neq 0$ (but the anticommutator term does vanish for an initial meter wave functions $\varphi_0(q)$ that is purely real).

These results mean that the weak value $S_w$ for the system's observable $S$ can be directly read-off from the (mean) position of the pointer variables $Q$ and $P$ after the post-selection operation. The results also generalize the results of the more brute-force method of sub-section V.A, the result of which within its limits is also confirmed.

It is also of some interest to calculate the second moment ${}_f\!<Q^2>$ of the pointer position and of its conjugate momentum. One finds

$$ {}_f\!<Q^2> \;=\; <Q^2>_0 + O(g^2)\,, $$

$$ {}_f\!<P^2> \;=\; <P^2>_0 + O(g^2)\,, $$

provided third-moment quantities like $<Q^2P>$ and $<P^3>$ vanish (which they do, *e.g.*, for a wave function $\varphi_0(q)$ symmetric around $q = 0$). The result means that the spread of the pointer states surviving the post-selection is approximately the same as the initial spread of the pointer states.

## V.E. *Application 2: Weak values for double qubits.*

The double qubit case of section IV.B requires a slightly different approach, since there I did not define any unitary operator $\mathcal{U}$. This case is especially interesting, since here all calculations can be made analytically without any approximations [17]; in the presentation here, however, I shall restrict myself to the weak measurement approximation.

The important starting point is the entangled wave function after the pre-measurement,

$$ \alpha\,|0>_{\mathbb{S}} \otimes (\,\cos\vartheta/2\;\;|0>_{\mathcal{M}} \;+\; \sin\vartheta/2\;\;|1>_{\mathcal{M}}) + \;\beta\,|1>_{\mathbb{S}} \otimes (\,\sin\vartheta/2\;\;|0>_{\mathcal{M}} + \cos\vartheta/2\;\;|1>_{\mathcal{M}}). $$

By inspection, one sees that a weak measurement is one in which the angular variable $\vartheta$ is near $\pi/2$. To exploit this fact, one puts $\vartheta = \pi/2 - 2\,\varepsilon$ and makes a Taylor series expansion in the (small) parameter $\varepsilon$.

I give the result only for one relevant entity, the Stokes parameter $S_1^{\mathcal{M}}$ for the meter. After some calculations one finds, neglecting terms of order $\varepsilon^3$,

$$ {}_f\!<S_1^{\mathcal{M}}> \;:= Tr\,(S_1^{\mathcal{M}}\,\mu_f) \;= \cos\vartheta\;\;\mathcal{R}e\,(\,<f\,|\,S_1^{\mathbb{S}}\;|s><s\,|\,f>)\,/\,prob\,(|f>|\,\tau_1) \approx $$

$$ \approx\;\varepsilon\;\mathcal{R}e\;\;(S_1^{\mathbb{S}})_w\,/\,\mathcal{D}_{dq}\;\;, $$

where, with the abbreviations $\alpha_f \;:=\; \alpha<f\,|\,0>$ and $\beta_f \;:=\; \beta<f\,|\,1>$ ,

$$ \mathcal{D}_{dq} \;:=\; 1 \;-\; \varepsilon^2\;\;\mathrm{Re}\,(\alpha_f\,\beta_f^{\,*})\,/\,|<f\,|s>|^2 \;\;, $$

and where

$$ <f\,|s> = \alpha_f + \beta_f $$

is assumed not to vanish.

### V.F. Application 3: Amplification with weak measurement.

From the treatment in section IV.C, the maximum amplification under the assumption made there – the essential points being the use of a von Neumann measurement on a qubit – was regulated by the overlap

$$r := \int dq\ \varphi_0(q-g)\ \varphi_0(q+g)^*,$$

in the sense that

$$|\,{}_f\!< Q >| \le g\,/\,\sqrt{1-r^2}\ .$$

In the weak measurement approximation,

$$r = 1 - \tfrac{1}{2}\ g^2 <P^2>_0 + O(g^3)$$

so that

$$\sqrt{1-r^2} \ \approx\ g\sqrt{< P^2 >_0}$$

and consequently

$${}_f\!<Q>_{max} \ \approx\ \frac{1}{\sqrt{< P^2 >_0}} \ \sim\ \sqrt{< Q^2 >_0}\ .$$

In other words, the weak measurement + post-selection procedure provides for a large amplification ${}_f\!<Q> / <Q>_1$ , the maximum of which is regulated by the spread of the initial meter wave function. However, the large amplification comes at a price: the corresponding value for the probability reads

$$prob\ (f\ |\ \tau_1)\ \approx\ g^2\ |\alpha_f|^2\ <P^2>_0\ \propto\ 1\,/\,({}_f\!<Q>_{max})^2\ ,$$

so a large amplification occurs only very rarely.

## VI. Testing the Leggett-Garg inequality with the double qubit.

As a non-trivial example of the use of the weak measurement concept, I will in this section apply it to one of the Leggett-Garg inequalities, derived in Appendix 4.

Bell derived his famous inequality from certain very general and 'reasonable' assumptions (locality, macro-realism). Quantum mechanics (QM) violates these inequalities, as do experiments. As a further probe on the relation of the micro world to the macro world, Leggett and Garg [11] derived another set of inequalities – to be abbreviated LGI – that again follow from some very general and reasonable assumptions: macroscopic realism and nonevasive measurability. The main difference to Bell's approach is that the LGI involve measurements at different times but on one and the same system; the LGI are also sometimes called 'Bell inequalities in time'. Appendix 4 gives a background to LGI. For some experiments on LGI, see [19, 23-25].

From Appendix 4, I take an LGI in the form applicable to a weak measurement + post-selection situation. The relevant LGI quantity is

$$< B > \; = \; < s|\, S \,|\, s > \; + \; |<f\,|s>|^2 \;(\mathcal{R}e\, S_w - 1\;)\,,$$

where as before

$$S_w \;=\; \frac{<f\,|S|\,s>}{<f\,|s>}$$

denotes the weak value of the observable $S$.

The LGI, assuming the eigenvalues of $S$ all lie in the interval [ –1, +1 ], reads

$$-\,3 \;\le\; < B > \;\le\; 1.$$

Let me now apply this LGI to the double qubit scheme of section IV.B.

Firstly, I note that the LGI is formulated only in terms of quantities referring to the object-system $\mathbb{S}$; however, the meter is so to speak in the background, needed at least for an experimental determination of $S_w$. Since only entities referring to the system are involved, I may use a simplified notation with no subscript $\mathbb{S}$ on the state vectors, *etc*; they now all refer to the object-system.

The initial state for the object-system is

$$|s > \;=\alpha\,|0 > \;+\; \beta\; |1 >,\;\; |\alpha|^2 \;+\; |\beta|^2 \;=1$$

The observable is taken to be

$$S = |0><0| \;-\; |1><1|,$$

(the Stokes' parameter $S_1$ for the photon case), and post-selection is made on

$$|f> \;=\; \cos \tfrac{1}{2}\varphi\,|0> \;+\sin \tfrac{1}{2}\varphi\;\; |1>.$$

It follows that

$$<s|\,S|\,s> \;=|\,\alpha|^2 \;-\; |\beta|^2\,,$$

$$<f\,|s> = \alpha\;\cos \tfrac{1}{2}\varphi \;+\; \beta\;\sin \tfrac{1}{2}\varphi\;,$$

and

$$|<f\,|s>|^2\;\;\mathcal{R}e\, S_w \;=\mathcal{R}e\,(\,<s\,|f><f\,|\,S|\,s>\,) = \;|\alpha\;\cos \tfrac{1}{2}\varphi|^2 \;-\; |\beta\;\sin \tfrac{1}{2}\varphi|^2\;.$$

The LGI entity $< B >$ under these particular circumstances then reads

$$< B > = |\alpha|^2 - |\beta|^2 + |\alpha \cos ½\varphi|^2 - |\beta \sin ½\varphi|^2 - |\alpha \cos ½\varphi + \beta \sin ½\varphi|^2$$

It may take values up to a maximum of $1\frac{1}{12}$, occuring for $\beta = -\frac{1}{6} = \cos \varphi$ , thus exceeding the LGI limit of +1.

In summary, the double qubit version of the weak value protocol affords an interesting testing ground for LGI. Concretely, it proves that QM violates the LGI. The experiments [19, 23-25] that test LGI in other ways show the same. Since all these approaches apply weak measurements, the Leggett-Garg assumption of non-invasive measurement (see Appendix 4) is (approximately) fulfilled. Consequently, the reason for violating LGI can be uniquely traced to the QM violation of the other main assumption in deriving LGI: quantum mechanics as well as nature do not respect the assumption of macroscopic definiteness.

## *VII. An experimental bouquet*

Several, often very ingenious, experiments have been performed utilizing the weak measurement + post-selection scheme. From the way they apply the scheme – and not primarily from the experimental methods used – they can be very roughly divided into two categories.

The first category comprises experiments that in one way or another test basic premises of the scheme or (some of) the proposals that the proponents of the scheme have suggested. They include tests of the so called Hardy's Paradox [26-28] and the Three-Box Paradox [7, 8, 29, 30] as well as some tests of the Leggett-Garg inequality [11, 19, 23-25]. The experimental methods used here are essentially based on the double-qubit scheme of section IV.C in its photonic polarization version, but the use of solid state devices [25] for investigating weak values may also be placed in this category. I comment very briefly on these experiments in the first sub-section below.

To the second category I count some very innovative experiments [31-33, 35 37,38], where the advantages of the weak measurement + post-selection scheme have brought really new insight. They will be treated in somewhat more detail, even if still sketchy, in the other subsections below. Needless to say, the choice of these particularly beautiful flowers in the experimental bouquet is highly subjective.

From the outset, let me also say that I will not be able to do justice to the often very intricate experimental techniques developed. I will merely focus on how the experiments in question throw light on the more theoretical and principal questions.

### *VII.A. Qubit measures qubit*

In most case, the experimental realization of the qubit-measures-qubit protocol of sub-sections IV.B and V.E has been through the use of entangled polarized photons [19]. Two polarized photons in the initial state are entangled, *e.g.*, to give the exclusive-OR gate of sub-section IV.B or some similar entanglement; this could be done by a technique with a set of polarizing beam splitters and so called half-wave plates [19]. The experiments also rely on techniques to measure the polarization of a photon without destroying it [38].

Here, I shall restrict myself to this very brief sketch of the principles; in section VIII.B below, I shall comment on their relevance for understanding what a weak value really means.

## *VII.B. Amplification*

We have seen in section IV.C how post-selection opens up the possibility of amplification, in particular if the measurement is weak (section V.E). In fact, the pioneering work [6] caught the interest of experimentalists. The theoretical ideas were for the first time implemented experimentally in a paper from 1991 by Richtie *et al* [31].

Their experiment can be characterized as an optical analog of a Stern-Gerlach set-up for electrons. They used a birefringent crystal to give different directions to the two linear-polarization components – to be identified with the degrees of freedom of the object-system in my general presentation – of a laser beam that passes through the crystal. But the separation between the two emerging beams is much too small to be detectable. So, with a second polarization analyzer the experimenters make a post-selection on the polarization of the beam. If this post-selected state is almost orthogonal to the initial state, the maximum mean value $_f{<}Q{>}_{\max}$ of the meter pointer variable – in this case the (transverse) position of the beam on a 'screen' (actually a photo detector) – will be clearly displaced from the beam position without the polarizer. This amplification effect thus arises from the fact that the expression for $_f{<}Q{>}$ is proportional to the weak value of the polarization, in particular to the (small) denominator $<f\,|s>$. All this is consistent: the imprecision $\Delta k_y = \sqrt{< P^2 >_0}$ in the transverse momentum of the incoming laser beam can be made very small, implying that the bound $_f{<}Q{>} \lesssim 1/\Delta k_y$ allows for large experimental values of $_f{<}Q{>}$.

A remark in passing: note that here and in the next experiments described, the 'meter' and the 'system' refer to two properties – a photon's (transverse) momentum and its polarization, respectively – of one and the same physical object.

A very similar idea lies behind the experiment by Hosten and Kwiat from 2008 [32]. They investigated the so called spin Hall effect for light, SHEL. This effect amounts to a (transverse) splitting of a well collimated laser beam into two parallel beams of different circular polarizations, when the beam under certain circumstances passes from one material to another with different refractive indices. The splitting effect is, however, tiny – a fraction of the wavelength of the light – and requires amplification. To generate the SHEL splitting, the experimenters sent a linearly polarized laser beam at an angle through a glass-air interface. To enhance the SHEL, the experimenters only analyzed outgoing photons polarized almost at $90^o$ with respect to the incoming beam (post-selection). In this way, the weak measurement + post-selection effect amplifies the original displacement, in the actual case by four orders of magnitude, corresponding to a sensitivity of ~ 0.1 nm. Again, this agrees with the general treatment in section IV.C and V.E.

Still another amplification effect exploiting the weak measurement + post-selection scheme was investigated by Dixon *et al* in 2009 [33]. They used a so called Sagnac interferometer. Very roughly, in their set-up light is sent into a channel formed as a square circuit. The light enters the circuit in one corner through a 50/50 polarizing beam splitter, thus, depending on the polarization, being directed into one of the two arms emanating from that corner. The beam then reflects off (one of the) fixed mirrors in the adjacent corners and off a tiltable mirror in the diagonally opposite corner. The light continues and reaches the other fixed mirror in the second adjacent corner, finally to make the full tour and return to the corner of the circuit from where it entered. Moreover, one may suitably insert into the circuit a polarization-dependent device that shifts the phase for light travelling the circuit in one direction but not in the other. In this way, the experimenters can vary the amount of light that leaves the device through the"bright port" (leaves the beam-splitter corner in the same direction as it entered) compared to that which leaves through the "dark port" (the orthogonal one). The experimentalists studied how a well-collimated beam passing through the device shows up at a detector 'screen' (actually a so called quadrant detector) placed behind the dark port. The game is to (weakly) entangle the direction of the beam (the meter) with the clockwise-counterclock-wise, polarization-dependent direction of the light through the device so that only a small amount comes out in the dark port (post-selection). Slightly tilting the tiltable mirror results in different deflection of the beam at the 'screen' behind the dark port. The weak measurement + post-selection scheme implies a large amplifying effect. In fact, in agreement with the theoretical predictions, the experimenters were able to detect an angular deflection of the mirror by ~½ picorad, corresponding to a linear displacement of the mirror by ~10 fm !

## *VII.C. Measuring the wave function*

It has always been assumed that a direct measurement of the wave function for a system is impossible. Only indirect methods, such as tomographic techniques [34], were thought to be feasible. Not so with weak values.

Here is the magic [35]. Suppose you want the $x$-space wave function $\psi(x) = \langle x|s\rangle$ for the state $|s\rangle$. Simple in theory! In the expression $S_w = \dfrac{\langle f\,|S|\,s\rangle}{\langle f\,|s\,\rangle}$ for the weak value, just choose $S$ as the projector $\Pi_x := |x\rangle\langle x|$ onto the position eigenstate $|x\rangle$. For the post-selected state choose the momentum eigenstate, $|f\rangle = |p\rangle$. Then

$$S_w = (\Pi_x)_w = \frac{\langle p\,|x\rangle\langle x|\,s\rangle}{\langle p\,|s\,\rangle} = \frac{exp(i\,p\,x)\quad \psi(x)}{\hat{\psi}(p)}, \quad (\hbar = 1)$$

with $\hat{\psi}(p)$ the momentum space wave function. Finally, choose $p = 0$ and you get

$$\psi(x) = k\,(\Pi_x)_w\ ,$$

with $k$ a constant that can be determined later from the normalization of $\psi(x)$[13].

What seems so easy from the theoretical point of view is a challenging task experimentally. But in a paper by Lundeen et al from June 2011 [35], the task has been accomplished, at least in the special case of the transverse wave function for a photon.

Very roughly, the procedure is the following. The experimenters generated photons of a well-defined wavelength by a process of parametric down-conversion. Other optical devices put the photons in the beam in the state whose wave function is to be determined. The weak measurement of the transverse position, $x$, of the photons is accomplished by a device that (slightly) changes the polarization at a particular position, $x$, in the beam, differently for different transverse positions of the beam. The post-selection on zero (transverse) momentum is done by sending the beam through a Fourier Transform lense and then selecting only those photons that arrive at a central point, equivalent to $p = 0$. The final step of reading the meter consists in measuring the polarization of these selected photons. Suitable polarization parameters constitute the pointer response in the weak measurement + post-selection scheme. From these parameters one can determine both the real and the imaginary part of the weak value of the transverse-space projection operator, alias the transverse wave function.

In this experiment, note that it is the spatial properties – the photon transverse coordinate $x$ – that is to be identified with the 'system' and the photon polarization with the 'meter'.

## *VII.D. Determining the trajectories in a two-slit experiment*

In the first course on quantum mechanics we all learn that, in a double slit experiment, it is impossible simultaneously to have an interference pattern on the screen and to decide through which slit the particle went. But this conclusion only follows if you consider projective, i.e., strong measurements. For weak measurements the unavoidable disturbances due to the measurement are minimized. The intriguing possibility opens up of observing both position (which slit the particle went through) and momentum (the interference pattern), if not for each individual particle so in the sense of determining the average momentum for those particles that arrive at a given position. This has indeed been shown to be feasible experimentally in a paper by Kocsis *et al* from June 2011 [37].

Single photons were prepared in a double-slit-like situation by an optical device containing a 50/50 beam splitter as an essential ingredient. After exiting the (equivalent of a) slit screen, the photons get polarized into a pure linear polarization state – the diagonal state $|D> =$
$= 1/\sqrt{2}\,(|H> + |V>)$ where $|H>$ ($|V>$) is the horizontal (vertical) polarization state, respectively. The weak measurement imparts a small phase shift to the photonic state, different for $|H>$ and $|V>$, and depending on the photons' transverse momentum. The post-selection is on the position of the photon at a ‘detection screen’ (actually, a CDD device) which registers the photons’ transverse position, while the circular polarization of these photons act as the pointer variable. The photons registered at this screen build up an interference pattern, showing that the weak measurement does not appreciably disturb them. In other words, the experimenters can simultaneously determine both the photons’ position and their momentum. By placing the ‘detection screen’ at different distances from the ‘slit screen’ the experimenters were able to map (at least statistically) the full trajectory landscape. In the experiment, the (equivalent of the) distance between the two slits is 4.7 mm, and the detection screen is placed at distances from approximately 3 m to 8 m from the detection screen. The result is shown in figure 5.

[13] In a recent paper [36], Lundeen and Bamber have even shown how to generalize this scheme to determine the density matrix for a mixed state.

## *VIII. Interpreting weak values, pro and con*

So far, I have kept to a rather operational approach. All what I have treated follows from standard quantum mechanics (QM). This includes all formulae in which the weak value $S_w$ occurs, where $S_w = \frac{<f\,|S|\,s>}{<f\,|s>}$ for an observable $S$ of a system, subject to a ‘pre-selection’ in the state $|s>$ and ‘post-selection’’ in a state $|f>$, with $<f\,|s> \neq 0$.

In this section I venture to move into the territory of interpretation: What does a weak value really ‘mean’? This is a mined territory; many physicists have expressed strong opinions one way or another; see, *e.g.*, [7, 8] and references therein, as well as [12].

As the inventors of the weak measurement + post-selection protocol, Aharonov, Vaidman and their collaborators are the foremost advocates of the novelty, usefulness, and even paradox-solving abilities, of the concept of weak value. And they have been very successful in their endeavor.

Of course, I shall not aim for a verdict in this slightly infected debate. What I will do is to present some arguments, both pro and con, that have some bearing on the issue. I shall not draw any final conclusion but leave a few question-marks to be straightened-out by further debate.

### *VIII.A. Basics*

The weak value is a formally well-defined concept. Its definition includes well-defined entities entirely referring to an object-system’s Hilbert space.

There has been some criticism focused on mathematical deficiencies [12]. True, in the presentations in the literature one may find some cavalier mathematics and ditto logic. The derivations presented in this paper, derivations that are maybe a little stricter than other ones, are of course still very cavalier from a mathematical point of view. I think, though, that they can be made mathematically strict with more carefully introduced assumptions and more attention to formal deduction; an effort in this direction is in [39]. In other words, I do not subscribe to an opinion that mathematical deficiencies should weaken the approach to weak measurements.

In arguing about the weak value, one should particularly bear in mind three circumstances.

Firstly, for a given system the weak value depends on three entities: the pre-selected state $|s>$, the observable $S$, and the post-selected state $|f>$. There is sometimes a tendency to forget, *e.g.*, that the dependence on the post-selected state is as important as the dependence on the pre-selected state.

Secondly, the weak value occurs in, and is measured by, the ancilla measurement scheme. It cannot be obtained in a (projective) measurement on the system alone. In particular, it cannot be derived from the ABL rule for ideal measurements.

Thirdly, the weak value bears some analogy to an ordinary mean value of the observable in question. But this analogy should not be driven too far. One sees in the literature [40] descriptions like "the weak value of an observable $S$ (is) the average of a sufficiently large number of identical …. weak measurements of $S$ …. " and " … the average for the sub-ensemble post-selected in the state ( $|f>$), i.e. the weak value, is given by ($S_w$) …. ". I do not consider these statements to be adequate enough descriptions of the actual way the weak value is arrived at in the ancilla scheme.

## *VIII.B. What can experiment tell?*

As is evident even from the short presentation in section VII, experimentalists have found several ingenious ways of using weak measurement to extract novel information from quantum mechanical measurements. The idea of interpreting a particular weak value as (essentially) the wave function for the system [35], and the use of delicately chosen post-selection to amplify an effect [31-33] stand out as particularly striking examples. In this sense, the introduction of the concept of weak value has meant a real novelty.

Other uses of weak measurements concern experimental tests of some of the paradoxes that Aharonov and collaborators have analyzed in terms of weak values [27,29]. I shall comment on the interpretation of these analyses below. Here, I just want to emphasize that the fact that the experimental results in these cases agree with the theoretical prediction cannot be taken as an argument for any interpretation of the weak values. In all these cases, the theoretical predictions follow from conventional quantum mechanics as presented in this article. What these experiments show is thus that QM indeed gives the correct description of their particular experiment. And this is no small feat! But as a matter of principle – not of experimental ingenuity, technique or of anything else related to experiments and experimentalists – what the experiments test is QM, nothing else.

## *VIII C. The 'meaning' of weak value*

The really touchy question concerns the very interpretation of a weak value. What property of the system under study does it really represent?

The arguments from the proponents (see [7] and references therein) have been multifarious and eloquently presented and defended. It will take me much too far to even list them. But a hopefully fair summary of the essential ingredients seems to be the following.

For a strong ancilla measurement (which, as I remarked in section III.B, reproduces a projective measurement) the meter reveals the conventional quantities – like (combinations of) eigenvalues and probabilities for them to occur – according to the usual rules in the theory-experiment interface of QM (see section II above). There is no reason, so I think the argument goes, that making measurements weak should change this basic property. Consequently, since the transition from strong to weak measurement is a continuous one, the weak value in itself also represents some kind of (static, *i.e.*, not referring to any transitions) property of the system. Vaidman [41] expresses this in his definition of an 'element of reality' thus:

> *If we are certain that a procedure for measuring a certain variable will lead to a definite shift of the unchanged probability distribution of the pointer, then there is an element of reality: the variable equal to this shift.*

But one may argue against this opinion. In the conventional interpretation of QM, the ancilla scheme constitutes a concrete realization of the measurement process for determining, e.g., the possible values of the observable under study. In other words, the ancilla scheme rather requires that one takes the strong interaction limit; anything else is only approximations to the basic interpretational rules of QM.

Against the interpretation of the weak value as an 'element of reality' also stands the immediate impression that the weak value contains a matrix element $< f \mid S \mid s >$ between different states. Such an entity is in conventional QM associated with a transition (probability) amplitude. And an amplitude, so the QM rules say, must be squared to give a physically meaningful quantity. This would contradict the view that the weak value unsquared represents something physically.

Also, one should realize that a weak value is a statistical entity the way it is measured; there is no pointer pointing to it. Rather, a full pointer position distribution is required to deduce it experimentally, related as it is to a mean value of a pointer variable.

## *VIII.D. The Three-Box Paradox and weak values for number operators*

Let me dwell on one particularly intriguing context, *viz.*, one where weak values have been invoked to throw light on some QM paradoxes. I think in particular of the so called 'Hardy's Paradox' [26-28] and of the 'Three-Box Paradox' [7, 29, 30]. Since the basic arguments to 'explain´ these paradoxes by weak values have, in fact, a similar structure, I shall even limit myself to looking at the Three-Box Paradox, easier to explain as it is.

The argument goes as follows.

Imagine a single QM particle in any one of three boxes *A*, *B* and *C*, and described by the initial ('pre-selected') state $|s\rangle = 1/\sqrt{3}\ (|A\rangle + |B\rangle + |C\rangle)$. Suppose further that the particle is later found in the ('post-selected') state $|f\rangle = 1/\sqrt{3}\ (|A\rangle + |B\rangle - |C\rangle)$. Moreover, consider an (intermediate in time) measurement of the projection operator $\Pi_A = |A\rangle\langle A|$; it is a number operator telling whether the particle is to be found in box *A* or not. Likewise, $\Pi_B = |B\rangle\langle B|$ is the number operator telling if the particle is found in box *B*, and similarly for *C*. One is interested in the probability $prob_A(\text{in } A)$ for finding the particle in box *A* when measuring $\Pi_A$, as well as the corresponding probabilities for *B* and *C*.

Consider first an ideal measurement of the respective projection/number operator. The ABL rule as presented in section II.C applies and gives

$$prob_A(\text{in } A) = 1 = prob_B(\text{in } B) \quad \text{while} \quad prob_C(\text{in } C) = 1/5$$

At first, there seems to be a paradox here. Not only does the total probability to find the particle in any box exceeds 1, it is with certainty – or at least with probability 1 – found both in box *A* and in box *B*. But the paradox disappears when one realizes that the results apply to different, projective measurements, which certainly cannot all be performed without each measurement heavily disturbing ('collapsing') the system and thereby creating totally new conditions for the next one.

Could a weak measurement come to the rescue? At least it does not collapse the system. In fact, nothing forbids us to do all three (weak) measurements of the number operators $\Pi_A$, $\Pi_B$ and $\Pi_C$ successively on the same pre-and post-selected states.

The corresponding weak values are

$$(\Pi_A)_w = 1 = (\Pi_B)_w \ .$$

On the other hand,

$$(\Pi_A)_w + (\Pi_B)_w + (\Pi_C)_w = (\ \Pi_A + \Pi_B + \Pi_C\ )_w = 1,$$

which together imply

$$(\Pi_C)_w = -1 \ !$$

Consequently, if one interprets also the weak value as a *bona fide* value of a number operator, one arrives at the mind-boggling result that there is minus one particle in box *C*!

Let me see how this stands further scrutiny.

The 'strong' values of a projection operator are its eigenvalues, 1 and 0. It is this that legitimizes the result of a measurement of a projection operator to be interpreted as the number of particles in the respective box, and its mean value as the probability of finding the particle in that box (*c.f.* section II.A). But how legitimate is it to interpret the weak value of a projection operator as 'the', or even 'a', particle number, or as a probability?

A clue to answering that question sits in the observation that a weak value depends not only on the pre-selected state, $|s\rangle$, but also on the post-selected one, $|f\rangle$. And by considering different combinations of the basis states $|A\rangle$, $|B\rangle$ and $|C\rangle$ for $|f\rangle$, one may get essentially any result for the weak value of a number operator. I devote Appendix 5 to a more in-depth analysis of this issue.

In conclusion, there is at least a question mark on how to interpret the weak value of a projection/number operator. This question mark propagates to the meaning of the weak value in general.

## *IX. Summary*

I have given a review of the very interesting and active field of weak measurements. In particular, I have focused on the way Aharonov, Vaidman and their collaborators have developed it. I have underlined the essential role played by post-selection in their methods as well as in general for amplification purposes. I have shown that the weak measurement approach requires no new features besides conventional quantum mechanics (QM) to arrive at the main results. I have also emphasized the influence the approach has had on experiments, triggering some ingenious experimental applications that relax some of the restrictions that are attached to the standard, projective ('strong') QM measurement scheme.

Where a question mark remains is on the interpretation of the weak value. As I have argued, it seems to me that the concrete interpretation made by Aharonov *et al* of the weak value, *i.e.*, essentially as a generalization of the ordinary mean value of an observable, is not without objections.

*Acknowledgement*

I thank S. Kröll, L. Sanders and B. Söderberg for comments on (part of) the manuscript, M.-E. Pistol for constructive suggestions, H.F. Hoffman for interesting discussions over the net and for drawing my attention to the work [48] by Kofman *et al*, and S Ashhab for a careful reading of the first version of my article.

# *Appendix 1. Some properties of measurement operators and effects, alias POVM*

The so called measurement operators are defined in terms of the pre-measurement unitary evolution operator $\mathcal{U}$ in sub-section III.C. They play a relatively minor role in my presentation. This is because I prefer to rely on more explicit expressions for $\mathcal{U}$, and to derive the results I want from the ensuing time-evolution of the density matrices. However, in more general treatments of the measurement process in quantum mechanics, they and the 'effects' – to be defined in point (vii) below – play a much more central role; see, *e.g.*, [5]. Therefore, it is of interest also here to list and derive some of their properties, which I do in this appendix, making heavy use of the notation and the results from section III.

## *1.A. Basic properties*

(i) A measurement operator is defined by $\Omega_k := < m_k \mid \mathcal{U} \mid m^{(0)}>$. It is an operator on the Hilbert space $\mathcal{H}_{\mathbb{S}}$ of the system.

(ii) Some elementary relations are

$$\Omega_k = < m_k \mid \mathcal{U} \mid m^{(0)}> = \sum_i < m_k \mid m^{(i)} > |s_i> < s_i| =$$

$$= \sum_i < m_k \mid m^{(i)} > \Pi_{s_i}$$

(iii) If one compares the expression

$$\sigma_1(\ |m_k) = \frac{1}{prob\ (m_k)}\ \Omega_k\ \sigma_0\ \Omega_k{}^\dagger$$

to the Lüders' rule

$$\sigma_1(\ |s_i) = \frac{\Pi_{s_i}\ \sigma_0\ \Pi_{s_i}}{prob\ (s_i \mid \sigma_0)}$$

in a projective measurement, one sees that the measurement operators $\Omega_k$ play the role of a kind of 'generalized projectors'. However, they do obey different rules from ordinary projectors.

(iv) Projectors obey

$$\Pi_{s_i}{}^\dagger = \Pi_{s_i}\ ,$$
$$\Pi_{s_i} \times \Pi_{s_j} = \delta_{i,j}\ \Pi_{s_i}\ .$$

For measurement operators such rules, in general, are not valid:

$$\Omega_k{}^\dagger = \sum_i < m^{(i)} \mid m_k > \Pi_{s_i} \neq \Omega_k\ ,$$

and

$$\Omega_k^\dagger \times \Omega_l = \Big( \sum_i < m^{(i)} \mid m_k > \Pi_{s_i} \Big)^\dagger \times \Big( \sum_j < m_l \mid m^{(j)} > \Pi_{s_j} \Big) =$$

$$= \sum_i < m^{(i)} \mid m_k > \Pi_{s_i} < m_l \mid m^{(i)} >$$

which in general has no simple relation to either $\Omega_k$ or $\Omega_l$ ; remember that the states $|m^{(i)} >$ in general do not form a complete set in $\mathcal{H}_{\mathcal{M}}$.

(v) Note however that

$$\Omega_k^\dagger \times \Omega_k = \sum_i < m^{(i)} \mid m_k > \Pi_{s_i} < m_k \mid m^{(i)} > = \sum_i |< m^{(i)} \mid m_k >|^2 \Pi_{s_i} ,$$

implying that

$$\sum_k \Omega_k^\dagger \times \Omega_k = \mathbb{I}_{\mathcal{S}} ,$$

with $\mathbb{I}_{\mathcal{S}}$ the unit operator in the system Hilbert space $\mathcal{H}_{\mathbb{S}}$; in deriving this relation I have used the facts that the basis states $|m_k >$ in the meter Hilbert space $\mathcal{H}_{\mathcal{M}}$ form a complete set, and that the states $|m^{(i)}>$ are normalized. This relation could be compared to the completeness relation $\sum_i \Pi_{s_i} = \mathbb{I}_{\mathcal{S}}$ for usual projectors.

(vi) The measurement operators $\Omega_k$ also enter into the expression for the probability *prob* $(m_k)$ to obtain the value $m_k$ :

$$Tr_{\mathbb{S}} ( \Omega_k \, \sigma_0 \, \Omega_k^\dagger ) = Tr_{\mathbb{S}} ( \Omega_k^\dagger \, \Omega_k \, \sigma_0 ) =$$

$$= \sum_j \Big( \sum_i \big| < m^{(i)} \, | m_k > \big|^2 <s_j \mid \Pi_{s_i} \, \sigma_0 \mid s_j> \Big) =$$

$$= \sum_i \big| < m^{(i)} \, | m_k > \big|^2 <s_i \mid \sigma_0 \mid s_i> = prob \, (m_k).$$

In fact, this is a consistency condition, since it is equivalent to the normalization condition $Tr_{\mathbb{S}} \, \sigma_1 ( \, | m_k) = 1$.

(vii) The product $\Omega_k{}^\dagger \Omega_k =: \mathcal{E}_k$ , appearing naturally in this expression for *prob* $(m_k)$, has many names. It is called “effect” or “probability operator”, sometimes alternatively “probability-operator-(or positive-operator-)-valued measure” (abbreviated POVM or POM). With this notation, the relation $\sum_k \Omega_k^\dagger \Omega_k = \mathbb{I}_{\mathcal{S}}$ derived in point (v) above reads

$$\sum_k \mathcal{E}_k = \mathbb{I}_{\mathcal{S}} ;$$

it is an expression of the fact that probabilities

$$prob \, (m_k) = Tr_{\mathbb{S}} \, (\mathcal{E}_k \, \sigma_0)$$

sum to unity.
Thus not only the measurement operators $\Omega_k$ but also the effects $\mathcal{E}_k$ have properties similar to projectors. The effects are Hermitian,

$$\mathcal{E}_k^\dagger = \mathcal{E}_k \ ,$$

but in general not idempotent:

$$\mathcal{E}_k \times \mathcal{E}_l \neq \delta_{k,l}\, \mathcal{E}_k \, .$$

(viii) A more realistic model of the ancilla scheme takes into account the fact that the measurement process could involve more degrees of freedom than the one, $S$ , to be measured for the system, and the pointer observable $M$. So let me suppose that the added degrees of freedom, referred to as $D$ , are described in a Hilbert space $\mathcal{H}_{\mathfrak{D}}$ with basis states $|\, d_r >$ , $r = 1, 2, \ldots, d_D = \dim(\mathcal{H}_{\mathfrak{D}})$. The initial state is $|d^{(0)}>$ . The pre-measurement prescription of section III.B is replaced by[14]

$$|s_i > \otimes\, |d^{(0)}> \otimes\, |m^{(0)}> \ \overset{\mathcal{U}}{\rightarrow}\ \ \mathcal{U}\,(\,|s_i > \otimes\, |d^{(0)}> \otimes\, |m^{(0)} >\,)\ =:\, |s_i > \otimes\, |d^{(i)}> \otimes\, |m^{(i)} > .$$

Thus, the measurement process is supposed to influence also the new degrees of freedom. The difference is that they are of no interest for that process and must consequently be summed over. In particular, when forming the conditional density matrix $\sigma_1\,(\,|m_k)$ after the read out of $M$ with the result $m_k$, one must trace out also the $D$ degrees of freedom (*c.f.* section III.C), resulting in

$$\sigma_1\,(\,|m_k) \ = \ Tr_{\mathcal{M},\mathcal{D}}\ \tau_1(\,|m_k) \ =$$

$$= \frac{1}{prob\ (m_k)}\ Tr_{\mathcal{M},\mathcal{D}}\ \{(\,\mathbb{1}_{\mathcal{S}} \otimes \mathbb{1}_{D} \otimes \Lambda_{m_k}\,)\,\tau_1\ \ (\mathbb{1}_{\mathcal{S}} \otimes\ \mathbb{1}_D \otimes \Lambda_{m_k}\,)\,\} =$$

$$= \frac{1}{prob\ (m_k)}\ \Sigma_r \,\big(\, < d_r \,|\otimes < m_k \,|\ \ \mathcal{U}\, |d^{(0)}> \otimes\, |m^{(0)}>\ \ \sigma_0$$

$$<m^{(0)}\,| \otimes\, < d^{(0)}\,|\ \mathcal{U}^\dagger\,|\ m_k > \otimes\ |\ d_r >\big) =$$

$$= \frac{1}{prob\ (m_k)}\ \ \Sigma_r\ \ \Omega_{k;r}\ \ \sigma_0\ \ \Omega_{k;r}^\dagger\ ,$$

with

$$\Omega_{k;r}\ := \ < d_r \,| \otimes < m_k \,|\ \ \mathcal{U}\, |d^{(0)}> \otimes\, |m^{(0)}> \ = \ \Sigma_i < m_k \,|\ m^{(i)} > < d_r \,|\ d^{(i)}> \ \prod_{s_i} \ \ .$$

The probability becomes

[14] One could think of even more general situations with the *D* degreees of freedom entangled with the system. However, this will not add anything substantially new, so I keep to the simple case presented here.

$$prob\,(m_k) = Tr_{\mathbb{S}}\,(\,\Sigma_r\ \Omega_{k;r}\ \sigma_0\ \Omega_{k;r}{}^\dagger\,) = \Sigma_r\ Tr_{\mathbb{S}}\,(\,\Omega_{k;r}{}^\dagger\,\Omega_{k;r}\ \sigma_0\,),$$

implying that the effects read

$$\mathcal{E}_k := \Sigma_r\,\Omega_{k;r}{}^\dagger\,\Omega_{k;r}\ .$$

The expressions for the relations between the density matrix $\sigma_1$ ( $|m_k)$, the measurement operators $\Omega_{k;r}$ , and the effects $\mathcal{E}_k$ derived here are the most general ones in the general theory of measurement (see ,*e.g.*, [5]).

## 1.B. Realization in the von Neumann protocol

The von Neumann protocol of sub-section IV.A lives in the $q$-basis. This means the following concrete realizations of the measurement operators

$$\Omega_k = \Sigma_i < m_k\,|\,m^{(i)} > \Pi_{s_i} \hookrightarrow \Omega_q = \Sigma_i\ \varphi_i(q)\,\Pi_{s_i} = \Sigma_i\ \varphi_0(q - g\,s_i)\,\Pi_{s_i} \quad .$$

For the effects

$$\mathcal{E}_k = \Sigma_i\ |< m^{(i)}\,|\,m_k >|^2\ \Pi_{s_i} \hookrightarrow \mathcal{E}_q = \Sigma_i\ |\ \varphi_0(q - g\,s_i)\,|^2\,\Pi_{s_i}\ .$$

## 1.B. Realization in the double qubit protocol

In the double qubit protocol of section IV.B, there are two measurement operators, $\Omega_0$ and $\Omega_1$. They take the form (recall what I said in footnote 9 concerning notation!)

$$\Omega_0 = {}_{\mathcal{M}}\!< 0|m^{(s_i=+1)} >_{\mathcal{M}}\ \Pi_{s_i=+1} + {}_{\mathcal{M}}\!< 0|m^{(s_i=-1)} >_{\mathcal{M}}\ \Pi_{s_i=-1} =$$

$$= c\ |0>_{\mathbb{S}\ \mathbb{S}}\!< 0\,| + s\ |1>_{\mathbb{S}\ \mathbb{S}}\!< 1\,| = \begin{pmatrix} c & 0 \\ 0 & s \end{pmatrix},$$

and

$$\Omega_1 = {}_{\mathcal{M}}\!< 1|m^{(s_i=+1)} >_{\mathcal{M}}\ \Pi_{s_i=+1} + {}_{\mathcal{M}}\!< 1|m^{(s_i=-1)} >_{\mathcal{M}}\ \Pi_{s_i=-1} =$$

$$= s\ |0>_{\mathbb{S}\ \mathbb{S}}\!< 0\,| + c\ |1>_{\mathbb{S}\ \mathbb{S}}\!< 1\,| = \begin{pmatrix} s & 0 \\ 0 & c \end{pmatrix}.$$

where I used the short-hand notation $c := \cos\vartheta/2$ and $s := \sin\vartheta/2$ , and where I also explicitly introduced the matrix representations.

The probability operators (or effects or POVMs) then become

$$\mathcal{E}_0 = \Omega_0^\dagger \text{ x } \Omega_0 = \begin{pmatrix} c^2 & 0 \\ 0 & s^2 \end{pmatrix},$$

$$\mathcal{E}_1 = \Omega_1^\dagger \text{ x } \Omega_1 = \begin{pmatrix} s^2 & 0 \\ 0 & c^2 \end{pmatrix}.$$

## *Appendix 2. Continuous measurements, the QM Zeno effect, and the master equation for an open system*

The treatment of weak measurement in the ancilla scheme in sub-section V.B is a convenient starting point for treating continuous measurements and for handling the influence of interaction with an 'environment' for an open system.

The transition to a continuous measurement is accomplished by performing a limit in which the time duration $\delta t_{\mathcal{U}}$ of the pre-measurement interaction between the system and the meter tends to zero. It is not straight-forward, though, just to take the limit $\delta t_{\mathcal{U}} \to 0$; some efforts are required in order to overcome a few stumbling blocks. In this appendix I give a broad outline of how this can be done.

### *2.A. The quantum Zeno effect*

I will consider a continuous measurement as the limit of a sequence of consecutive measurements, carried out at regular time intervals of a duration $\delta t_{\mathcal{U}}$ , which tend to zero. Thus, I model it by assuming the system to be subjected to *n* consecutive, identical measurements in time intervals between fixed times $t = t_0$ and $t = t_f$ (= $t_n$) as illustrated in figure 6. The difference $t_r - t_{r\text{-}1}$ between any two consecutive times is assumed to tend to zero as $n \to\infty$. Indeed, I assume each pre-measurement $\mathcal{U} = \exp(-i\,\gamma\, S \otimes P\, \delta t_{\mathcal{U}})$ to be of duration $\delta t_{\mathcal{U}}$ and to occur in the time interval between the indicated times so that $t_r - t_{r\text{-}1} = \delta t_{\mathcal{U}}$, $r = 1,2, \ldots n$. The limit to be taken is then $\delta t_{\mathcal{U}} \to 0$, $n \to\infty$ but such that $n \text{ x } \delta t_{\mathcal{U}} = t_f - t_0$ remains finite and non-vanishing. Consequently, a weak interaction approximation for $\mathcal{U}$ applies.

It follows from section III.D that the system density matrix $\sigma_n$ , after *n* such identical, consecutive measurements, reads

$$\sigma_n = \Sigma_{i,j} \; \Pi_{s_i} \,\sigma_0\, \Pi_{s_j} \,(< m^{(j)}|m^{(i)} >)^n .$$

Here, in the approximation outlined above,

$$<m^{(j)}|m^{(i)}> \;\approx\; 1 - \tfrac{1}{2}\gamma^2\ \delta t_{\mathcal{U}}{}^2 <N^2>_0\ (s_i - s_j)^2 =:\ 1 - \delta t_{\mathcal{U}}{}^2 A_{ij}\ ,$$

with for simplicity an obvious abbreviation.

Now, assume all entities except $\delta t_{\mathcal{U}}$ (and $n$) to remain finite and non-vanishing. Then

$$\lim_{n\to\infty}(<m^{(j)}|m^{(i)}>)^n = \lim_{n\to\infty}(1 - \delta t_{\mathcal{U}}{}^2 A_{ij})^n =$$

$$= \lim_{n\to\infty}(1 - \tfrac{1}{n}\ \delta t_{\mathcal{U}}\, t_f\, A_{ij})^n = \lim_{\delta t_{\mathcal{U}}\to 0}\ exp\,(-\delta t_{\mathcal{U}}\, t_f\, A_{ij}) = 1.$$

This is the quantum Zeno effect (see, *e.g.*, [42]): performing multiple identical measurements in short time intervals leave the system so to speak frozen in its initial state,

$$\lim_{n\to\infty}\sigma_n = \Sigma_{i,j}\ \Pi_{s_i}\,\sigma_0\,\Pi_{s_j} = \sigma_0\ ,$$

and no changes seem possible in such a continuous measurement.

## *2.B. Open system with smoothened environmental interaction*

There is, however, a way out: some other factor besides $\delta t_{\mathcal{U}}$ in the expression for $\mathcal{U}$ could also vary with time. For example, one could assume the coupling 'constant' $\gamma$ to vary. But a more natural and conventional choice is to let the spread $<N^2>_0$ in the meter variable be such that $<N^2>_0\ \delta t_{\mathcal{U}}$ tends to a finite value for $\delta t_{\mathcal{U}} \to 0$. In case $N = P$, as in the von Neumann protocol, this would mean a momentum spread for the meter $\Delta P_0{}^2 \propto 1/\ \delta t_{\mathcal{U}}$ so that the conjugate variable $Q$ would have a spread $\Delta Q_0{}^2 \propto \delta t_{\mathcal{U}}$ i.e., a more and more precise location of the pointer variable in the initial meter state the shorter the time duration gets. Anyhow, it is an assumption of this kind that is necessary for avoiding the QM Zeno effect [43, 44][15].

Before I enter on the nitty-gritty of the limiting procedure, let me widen the scope slightly. I shall not primarily consider any particular observable $S$ of the system. Instead, I shall be interested in the evolution of the system, in terms of its density matrix, under a continuous monitoring by an 'environment', *i.e.*, essentially anything that gently influences the system. I shall model this monitoring by assuming the interaction Hamiltonian $H_{\text{int}}$ to be of the form

$$H_{\text{int}} = \gamma\ \Sigma_a\ (\,T_a \otimes N_a\,)$$

with the operators $T_a$ acting in the Hilbert space $\mathcal{H}_{\mathbb{S}}$ of the system and $N_a$ in the meter Hilbert space $\mathcal{H}_{\mathcal{M}}$ . In fact, this is about the most general interaction Hamiltonian one may have for two interacting systems. I shall furthermore assume that all the operators $N_a$ have vanishing first moments in the initial state of the meter: $<N_a>_0\ = 0,\ \forall\ a$. The second moments shall obey $<N_a\, N_b>_0\ = \delta_{a,b} <N_a{}^2>_0\ ,\ \forall\ a,b$, *i.e.*, there should be no correlation between the different $N_a$ at this level.

---

[15] The usual procedure [43, 44] is to let the effective coupling constant $g = \gamma\ \delta t_{\mathrm{U}}$ be a real constant and to let $\Delta Q_0{}^2 \propto 1/\ \delta t_{\mathcal{U}}$ . One may convince oneself that the two procedures are equivalent. *C.f.*, footnotes 8, 11, 12.

Invoking some of the results derived in section V.B , assuming as before that the intrinsic Hamiltonians $H_{\mathbb{S}}$ and $H_{\mathcal{M}}$ vanish, and using the short-hand notation $\mathcal{X} = \int dt\, H_{\text{int}}$ , one finds for the last step in the time-series in figure 6 that

$$\sigma(t_f) := \sigma_n \approx \sigma_{n-1} + Tr_{\mathcal{M}}\big( i\ [\ \tau_{n-1}, \mathcal{X}\ ] - ½\,[\,[\tau_{n-1}, \mathcal{X}\ ], \mathcal{X}\ ]\ \big).$$

One more assumption is needed before I can get to the final result: The total system density matrix $\tau_r$ at any time $t_r$ is assumed to be up-dated at each measurement, so that $\tau_r = \sigma_r \times \mu_0$ with one and the same $\mu_0$ at all times. The motivation is that a new, but identical, meter is involved at each time.

Putting all pieces together, and using the notation $\sigma(t_f - \delta t_{\mathcal{U}}) := \sigma_{n-1}$ , results in

$$\sigma(t_f) - \sigma(t_f - \delta t_{\mathcal{U}}) \approx -\ ½\ \delta t_{\mathcal{U}}^{\ 2}\ \gamma^2\ \Sigma_a <N_a^2>_0 \big[\ [\ \sigma(t_f - \delta t_{\mathcal{U}}),\ T_a\ ]\ , T_a\ \big]\ .$$

Note in particular that there is no term linear in $\gamma$, since I assumed all $<N_a>_0$ to vanish. The last step consists in invoking

$$½\ \gamma^2 <N_a^2>_0\ \delta t_{\mathcal{U}} \rightarrow\ \eta_a^{\ 2} \neq\ 0 \text{ as } \delta t_{\mathcal{U}} \rightarrow 0$$

to derive , in the limit $\delta t_{\mathcal{U}} \rightarrow 0$ ,

$$\frac{\partial\sigma}{\partial t}(t_f) = -\ \Sigma_a\ \eta_a^{\ 2}\,\big[\ [\sigma(t_f),\ T_a\ ]\ , T_a\ \big]\ .$$

This is the so called 'master equation' in the very general 'Lindblad form' for a system in touch with an environment. At least almost. What remains is to generalize it to the case of a non-vanishing intrinsic Hamiltonian $H_{\mathbb{S}}$. This is easily done. So here is the very final result of this subsection (the time argument in the density matrix is now suppressed)

$$\frac{\partial\sigma}{\partial t} = i\,[\sigma,\ H_{\mathbb{S}}] -\ \Sigma_a\ \eta_a^{\ 2}\,[\ [\sigma,\ T_a\ ]\ , T_a\ ]\ ,$$

as usual in units such that $\hbar = 1$.

One may question some of the assumptions made in deriving this master equation. Indeed, other less pedestrian methods must be applied in a stricter derivation. The poor man's derivation presented here may anyhow be motivated by its comparative simplicity. See, *e.g.*, [5, 45, 46] for more basic approaches.

I shall not dwell upon applications of the master equation but refer to the literature. Let me only mention that one important such applications is the study of decoherence due to interaction with the environment [5, 46].

# *Appendix 3. Can the spin of a spin-½ particle be 100?*

The seminal paper [6] with the provocative title *"How the Result of a Measurement of a Component of the Spin of a Spin-½ Particle can Turn Out to be 100"* from 1988 by Aharonov, Albert and Vaidman was the spark that triggered the interest in weak measurements[16]. In the context of a Stern-Gerlach set-up, the authors showed how the weak value of a spin component could be measured and drew the conclusion that the spin of the particle could exceed 100.

In the present appendix, I shall investigate this claim from the purely formal point of view, leaving the interpretational aspect – what the weak value could really mean – to section VIII of the main text. Similar analysis can be found in, *e.g.*, [20-22].

The question to be considered in this appendix is this: How large can the weak value $S_w = \frac{<f\,|S|\,s>}{<f\,|s>}$ of an observable $S$ be? What does the formalism developed in this article have to say about large $S_w$?

It is clear that the numerator $< f\,|S|\,s >$ in the expression for the weak value is bounded provided, as I assume throughout, that the Hilbert space $\mathcal{H}_S$ of the system is of finite dimension. Consequently, a large $S_w$ requires a small denominator $< f\,|s >$. More precisely, granted that $< f\,|S|\,s > \neq 0$ – I will make this assumption throughout this appendix – the (absolute value of the) weak value $S_w$ can be as large as one likes by choosing a small value of (the absolute value of) $< f\,|s >$. In particular, $S_w$ may take values (far) outside the spectrum of $S$, i.e., the bounded interval defined by the eigenvalues of the operator $S$. Such values of $S_w$ have been baptized 'strange' values. In fact, I see nothing strange with them at all; they are just a consequence of the definition of $S_w$ and of the freedom to choose the post-selected state $|\,f >$ with respect to the pre-selected state $|s >$ so that $< f\,|s >$ becomes small.

The issue will be more relevant in case one interpret the weak value $S_w$ as a proper quality of the system: whether the fact that the pointer position taking large values – corresponding to large values of the weak value like that for the spin – really means that the spin is large is a matter of the meaning one attaches to the weak value. I have treated that interpretational question in section VIII.

An interesting question could, however, be how a large value of $S_w$ manifests itself in the measurement situation: how will the ancilla be affected by a large $S_w$?

The answer to that question sits in my treatment in sections IV.C and V.E of the amplification effect that are possible with post-selection in the ancilla scheme. True, I focused there entirely on the dependence of ancilla variables on $< f\,|s >$. But the main results will be the same if one considers the dependence on $S_w$ instead.

In the first special cases I dealt with, a von Neumann measurement of a qubit, a large value of the mean pointer variable ${}_f\!<Q>$ could be achieved by choosing a large value of $S_w$ (actually a small value of $< f\,|s >$). (Caveat: for $< f\,|s >$ really tending to zero, i.e., $S_w \to \infty$, one has in fact $_f\!<\!Q\!> \to 0$; in the notation of section IV.C, this occurs, however, only for a very narrow range, $|< f\,|s >| \leq \alpha_f \sqrt{< P^2 >_0}$, around zero.) So in that case 'strange' values of $S_w$ will have large effects on the ancilla. But for the double qubit example, irrespective of how large one makes the weak value $S_w$ of the system's Stokes' parameter, the corresponding ancilla variable stays within its 'non-strange' range.

[16] It is interesting to note that the paper was submitted June 30, 1987 but not published until April 4, 1988.

## *Appendix 4. The Leggett-Garg inequality, LGI*

From a few reasonable assumptions on what characterizes the macro world, Leggett and Garg [11] were able to deduce inequalities involving (in general different) measurements performed at successive times but on one and the same object-system. These inequalities are then generalized to the micro domain. In this appendix, I sketch the derivation of (one of) these inequalities, the one used in section VI.

The assumptions invoked by Leggett and Garg are, slightly reformulated,

*Macroscopic realism* (MAR): A macroscopic system will at all times be in one or the other of the states available to it.

*Noninvasive detection* (NID): It is possible to determine the state of the system with arbitrary small perturbations on its subsequent dynamics.

The set-up for LGI is described in figure 7. Three successive measurements of the observables $A_0$, $A_1$ and $A_2$ are performed. (For purely simplifying purposes, I shall assume that there is no time evolution in between the appropriate measurements.)

Consider first how the argument based on macroscopic reasoning goes. Suppose the result of a measurement of the observable $A_i$ is $a_i$ , $i$ = 0,1,2. Suppose also that all the $a_i$ lie in the interval [–1, +1]. Leggett and Garg then consider the combination

$$B := a_0\, a_1 + a_1\, a_2 - a_2 a_0$$

It is straight-forward to show the inequality

$$-\,3 \;\leq B \leq\; 1\,.$$

Averaging $B$ over the probability distributions for obtaining the respective values does not destroy the inequality – there are some subtleties involved, and one must of course use the fact that probabilities lie between 0 and +1 as well as the assumptions MAR and NID in the derivation – so with

$$< B > \;:=\; < A_0\, A_1 > + < A_1\, A_2 > - < A_2\, A_0 >,$$

(bra-kets < .. > denote averages over the respective probability distributions), one arrives at

$$-3 \leq \langle B \rangle \leq 1.$$

This inequality is then taken over to the micro domain by interpreting the averages as the QM expectation values in an initial state prepared before the measurement of $A_0$ takes place. It constitutes (one of) the Leggett-Garg inequality (-ties).

The LGI are aptly suited for studies using weak measurement: For a weak measurement, the condition NID of noninvasive measurement can be considered (approximately) fulfilled. Therefore, with weak measurement, LGI effectively only tests the macrorealism assumption, MAR. One may even relax a bit on the requirement of weak measurements. In fact, the main steps in the derivation of LGI really only require the middle measurement, that of $A_1$, to be weak. Both initially and finally, for $A_0$ respectively $A_2$, the measurement could be projective.

The actual applications [19, 23-25] even choose $A_0 = |s\rangle\langle s|$, the projection operator on the initial ('pre-selected') state. The Leggett-Garg entity $\langle B \rangle$ then reads

$$\langle B \rangle = \langle s|A_1|s\rangle + \langle s|A_1 A_2|s\rangle - \langle s|A_2|s\rangle ,$$

where, as remarked, $A_2$ may be measured projectively and only $A_1$ is required to be measured weakly. In the experimental realizations, different choices for $A_1$ and $A_2$ have been investigated [19-22].

One could fit the LGI even more closely to the theme of this article by considering a weak measurement + post-selection situation. By this I mean making the substitutions

$$A_0 \hookrightarrow \sigma_0 = |s\rangle\langle s|$$

$$A_1 \hookrightarrow S ,$$

$$A_2 \hookrightarrow |f\rangle\langle f| ,$$

which result in

$$\langle B \rangle = \langle s|S|s\rangle + \langle s|S|f\rangle\langle f|s\rangle - \langle s|f\rangle\langle f|s\rangle =$$

$$= \langle s|S|s\rangle + |\langle f|s\rangle|^2 \left(\frac{\langle s|S|f\rangle}{\langle f|s\rangle} - 1\right) , \quad \langle f|s\rangle \neq 0$$

This expression involves the weak value $S_w = \dfrac{\langle s|S|f\rangle}{\langle f|s\rangle}$ of the observable $S$.

Note, however, that there is an inherent order ambiguity in translating a classical mean value of a product of two observables into a QM expectation value when the two corresponding operators do not commute. In the last example, this is reflected in the fact that $S_w$, a complex number in general, appears in $\langle B \rangle$. I propose to solve the ambiguity by replacing an operator product, say $A_1 A_2$, by the symmetrized one, $\frac{1}{2}(A_1 A_2 + A_2 A_1)$.

Then, provided all the eigenvalues $s_i$ of $S$ lie in the interval $[-1, +1]$, the final expression for the Leggett-Garg inequality, LGI, becomes in this case

LGI: $-3 \leq \langle B \rangle \leq 1$ ,

with

$$\langle B \rangle = \langle s|\, S|\, s \rangle + |\langle f\,|s \rangle|^2 \left( \mathcal{R}e \, \frac{\langle s\,|S|\, f\rangle}{\langle f\,|s \rangle} - 1 \right).$$

I devote section VI of the main text to an analysis of this particular version of LGI in the double qubit set-up.

Summing up, the weak ancilla protocol has the advantage in connection with LGI that it obeys the non-invasive detection assumption, NID, in the limit of a very weak pre-measurement interaction, $g \to 0$. In particular, any violation of the LGI can then be blamed on the violation of the macroscopic realism assumption, MAR.

## *Appendix 5. Analysis of the Three-Box Paradox*

This appendix looks in more detail on some aspects of the Three-Box Paradox as presented in section VIII.D. I closely follow the treatment in [47]. The goal is to give a model of a physical mechanism that could reveal some features in the measurement of the projection/number operators involved.

Suppose we want to check whether there is a particle in box $C$, the one that had a weak value number operator equal $-1$ in the setting with a pre-selected state $|s\rangle = 1/\sqrt{3}\ (|A\rangle + |B\rangle + |C\rangle)$ and a post-selected one $|f\rangle = 1/\sqrt{3}\ (|A\rangle + |B\rangle - |C\rangle)$. Following [47], assume the particle one looks for is charged. One may then check its presence by shooting other charged (test) particles pass the box and see if they are deflected by the electric field of the hidden particle. Interpreted in the language I have used, the test particle is the meter and the hidden particle the system. The system observable to look for is its number operator $\Pi_C$. Furthermore, concentrate for simplicity only on (one of) the transverse degree(-s) of freedom, the transverse momentum $p$ of the meter, which is then chosen as the pointer variable (*c.f.* section III). A suitable interaction Hamiltonian could then be [47]

$$H_{\text{int}} = \gamma\ \Pi_C \otimes X\ ,$$

where $X$ is the (transverse) position of the meter particle; in effect this is a von Neumann protocol but with momentum $P$ and position $X$ ( $= Q$ ) interchanged. So I can take the relevant formulae from sections IV.A and V.D. In particular, the meter wave function $\varphi_1(p)$, after the particle has passed the box $C$ – *i.e.*, after the pre-measurement – is given by $\varphi_0(p - \langle s|\, \Pi_C\, |s\rangle)$ in the weak measurement approximation. (For simplicity, I here set the effective coupling constant $g = 1$; for those who do not like setting a small parameter = 1, just think of dividing suitable quantities by $g$ and use these ‘renormalized’ values.) In fact, with the choice made for $|s\rangle$, one has $\langle s|\, \Pi_C\, |s\rangle = 1/3$.

To study the effect of post-selection, let me assume a more general post-selected state

$$| f(\Theta)> \;:=\; \sin\Theta \;\; 1/\sqrt{2} \;\; (| A > + |B >) \; + \cos\Theta \;\; |C > \; ,$$

parameterized by an angle $\Theta$ with $0 \le \Theta \le \pi$. The pre-selected state is as before

$$|s > = 1/\sqrt{3} \;\; (| A > + |B > + |C >) \, .$$

The relevant entities are

$$< f(\Theta)|s > \;=\; 1/\sqrt{3} \;\; (\sqrt{2} \; \sin\Theta + \cos\Theta)$$

and

$$< f(\Theta)| \, \Pi_C \, |s > \; = 1/\sqrt{3} \;\; \cos\Theta \, .$$

So the weak value of the box particle number operator becomes

$$(\Pi_C)_w \;=\; 1 \, / \, (\sqrt{2} \;\; \mathrm{tg}\,\Theta \; + 1 \;) \;\; .$$

In agreement with the general scheme of section V.D – remember the interchange $P \leftrightarrows Q = X$ and that I have put $g = 1$ – this is then also the mean value $_f\!< P >$ of the pointer/meter particle transverse momentum after the post-selection of $| f(\Theta)>$ .

From these formulae – I have also illustrated the general situation in figure 8 – I conclude that the post-selected state $| f(\Theta)>$ acts as a kind of filter that 'selects' the pointer mean value $_f\!< P >$ after the post-selection. In particular, by varying $\Theta$ one may obtain any (real) value of $_f\!< P >$ and thus of the weak value $(\Pi_C)_w$ of the number operator.

This dependence on the post-selected state destroys, in my view, any interpretation of $(\Pi_C)_w$ as an inherent property of the particle in box $C$. In other words, trying to measure the content of the box with the proposed method will fail: what you get – the value of $(\Pi_C)_w$ – depends on what you select – the state $|f(\Theta)>$.

I add a few further remarks.

The definition of a weak value, like $(\Pi_C)_w$ , does not preclude it from taking complex values. This adds further doubts to the interpretation of a weak value as a property of the system.

There has been much discussion of the fact that the weak value of an observable may take so called strange value (*c.f.* Appendix 3), *i.e.*, values that lie outside the spectrum of eigenvalues of the corresponding operator. It is clear from figure 8, if not before, that strange values of, in this case, the number operator, occur due to suitable post-selection and to the large spread of the pointer position variable in the initial state. The figure also show that these values occur in the tails of the distribution, and therefore with low probability.

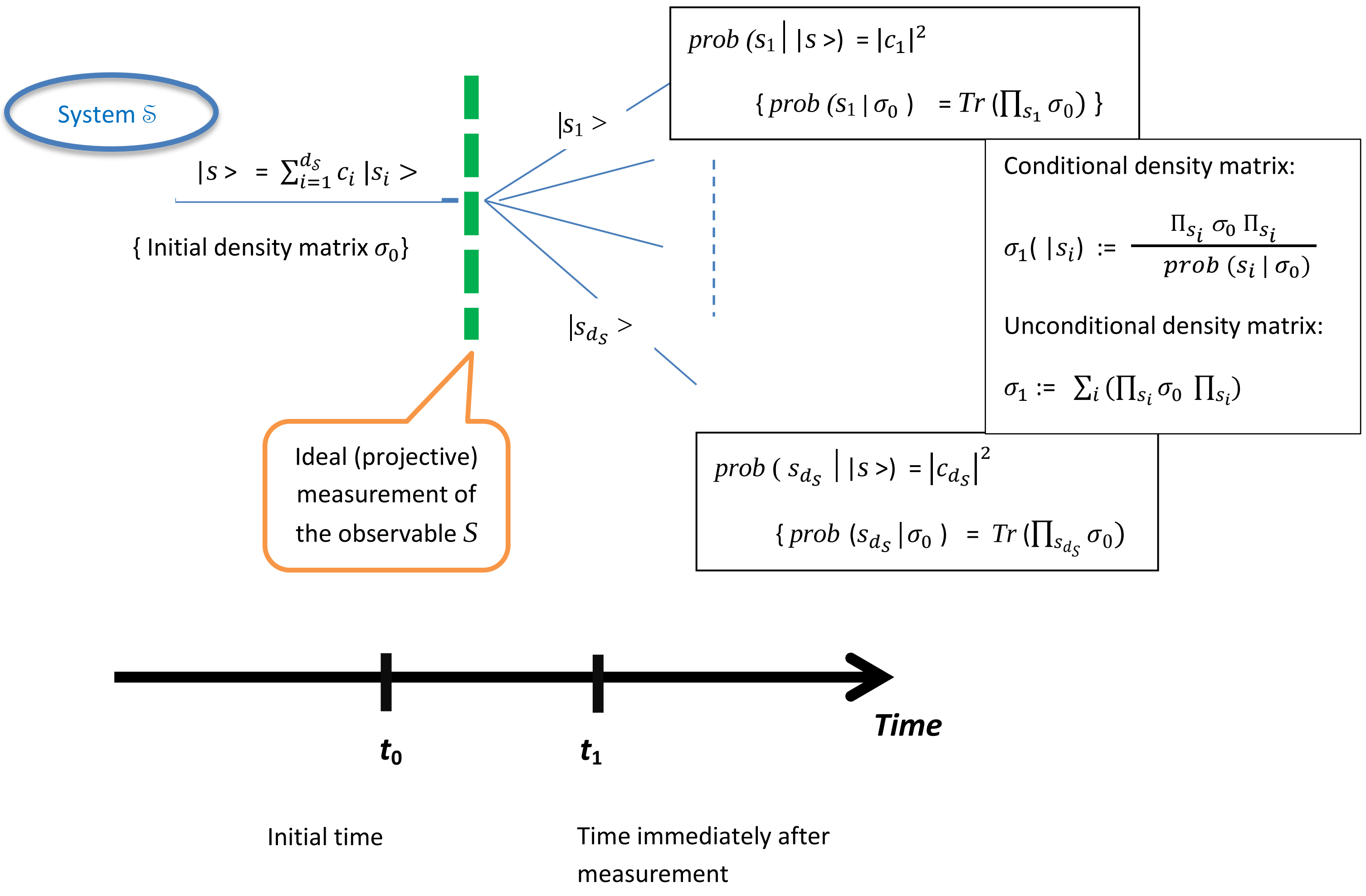

Figure 1. A diagrammatic representation of an ideal (or projective), non-destructive measurement, which also summarizes the notations used. The measurement is thought to occur in a definite time sequence (for simplicity, I assume that there is no 'internal' time evolution of the system, *i .e.*, its intrinsic Hamiltonian is assumed to be zero) : preparation (or 'pre-selection') at a time $t_0$ of an initial state $|s>$ (or initial density matrix $\sigma_0$), followed by a projective measurement of the observable $S$, resulting in a projection onto an eigenstate of the corresponding operator $\hat{S}$ or, if the result of the measurement is not registered (no 'read-out'), in an unconditional density matrix.

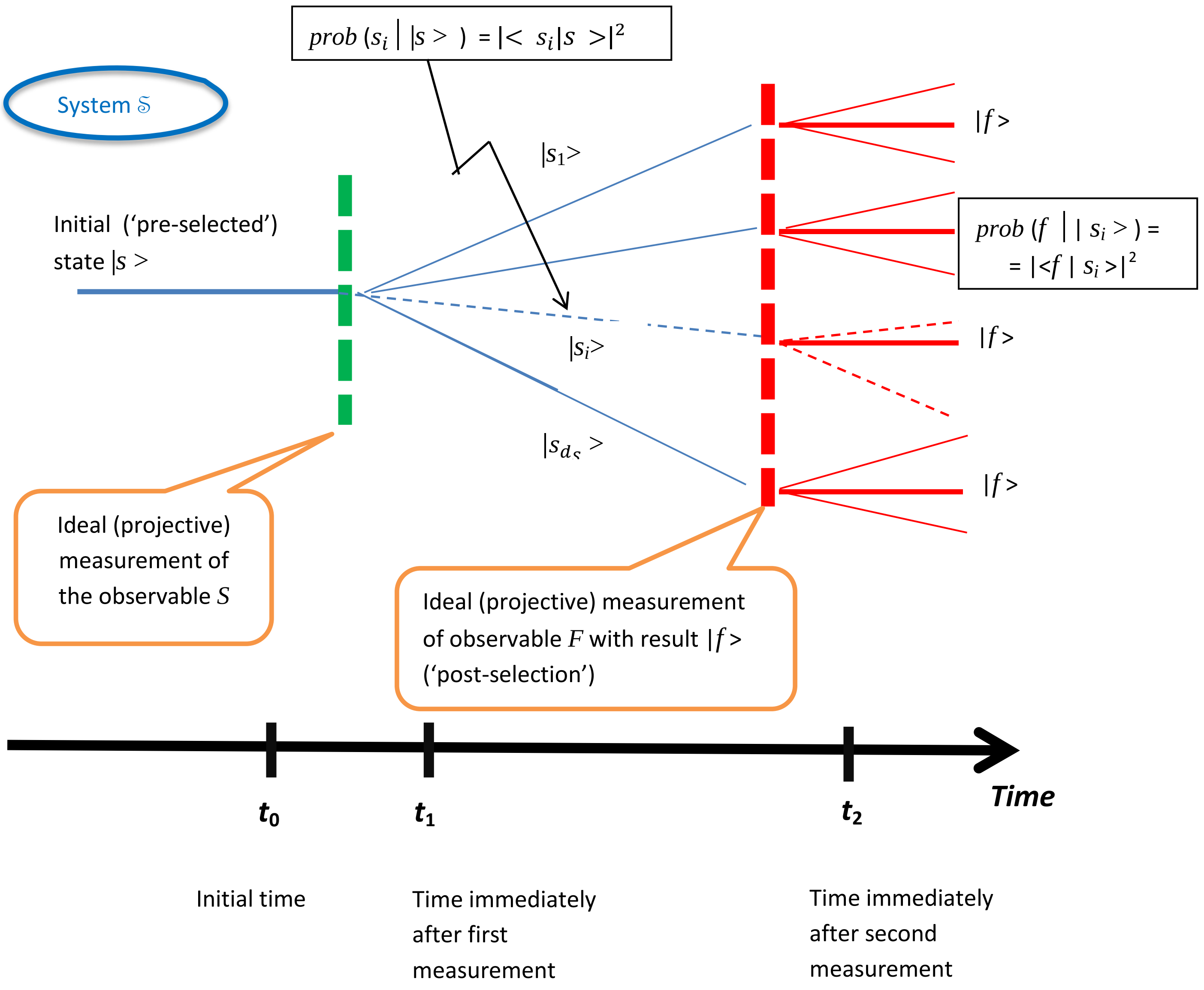

Figure 2. Illustrating the procedure behind the so called ABL rule [14]. With non-destructive measurements, one my envisage a situation where the system $\mathbb{S}$ is prepared (‘pre-selected’) in state $|s>$ (assumed for simplicity to be a pure state; the argument can equally well be carried through if the pre-selected state is a mixture), then subjected to a first measurement of the observable $S$, followed by a second measurement, now of another observable $F$. Of all the possible results of the this second measurement, only those which give rise to one particular ‘post-selected’ eigenstate $|f>$ of $F$ are kept in the ABL rule.

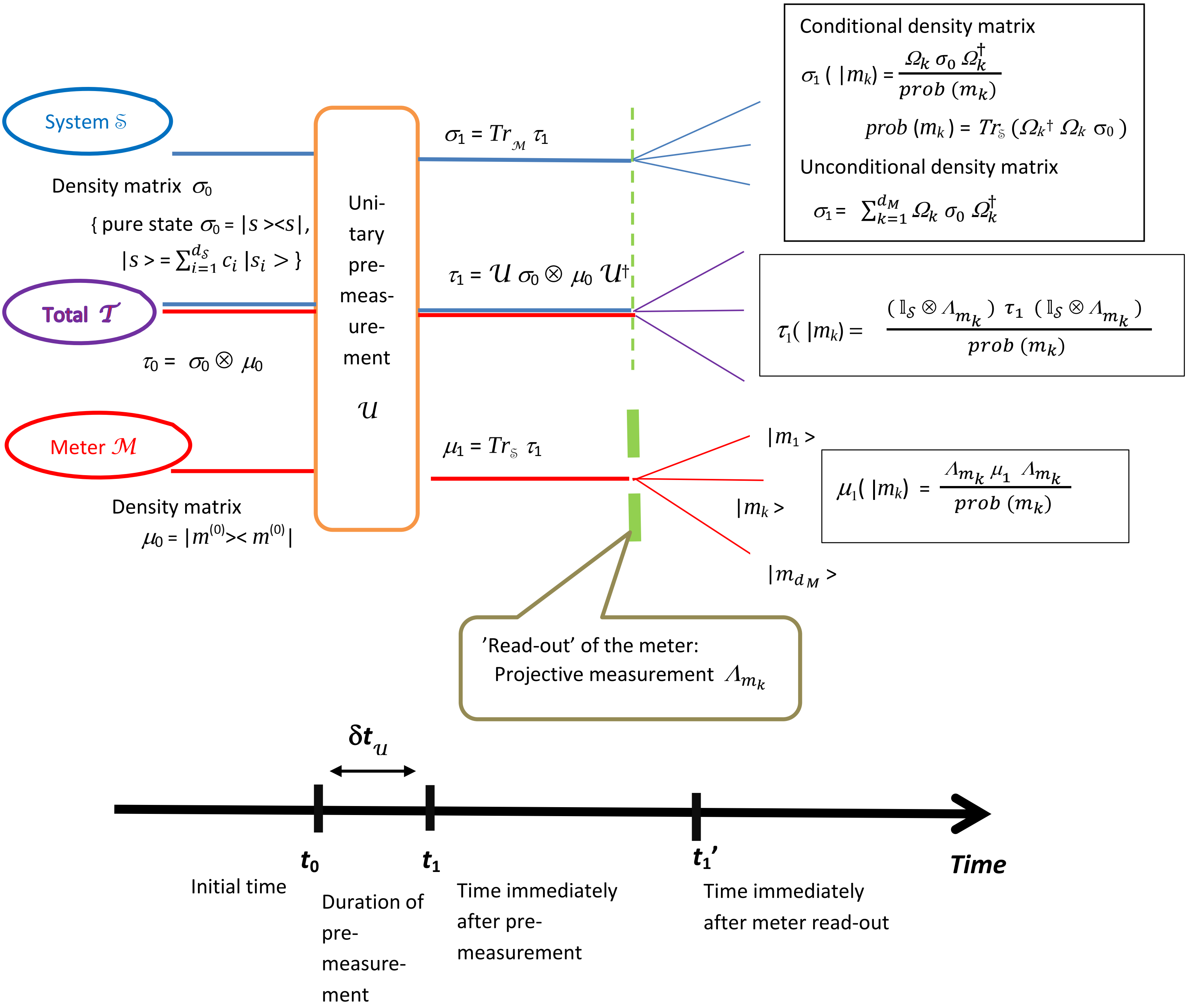

Figure 3. A diagrammatic representation of the indirect (or ancilla) measurement scheme. The object (or system) $\mathbb{S}$ and the measuring device ('the meter') $\mathcal{M}$ form together the total system $\mathcal{T}$. The scheme supposes that the system and the meter are prepared in (non-entangled) states $\sigma_0$ resp $\mu_0$ , so that the initial total state is $\tau_0 = \sigma_0 \otimes \mu_0$ (or, if pure states, $|s> \otimes |m^{(0)}>$). For simplicity, I assume throughout that there is no internal time evolution of the system and the meter, *i.e.*, that their respective intrinsic Hamiltonians vanish. The system and the meter then interact during a time $\delta t_{\mathcal{U}}$ via a (non-destructive) unitary interaction $\mathcal{U}$, which entangles them in a characteristic way encoded in the total density matrix $\tau_1$. After this pre-measurement, the meter is subjected to a 'read-out', a projective measurement, which produces one of the eigenstates $|m_k>$ of the 'pointer operator' $M$ which characterizes the meter in its Hilbert space $\mathcal{H}_{\mathcal{M}}$. Due to the fact that the system $\mathbb{S}$ is entangled with the meter, this measurement also influences the system $\mathbb{S}$. One can choose to neglect the result of the meter reading, in which case the system is finally described by the unconditional matrix $\sigma_1$. But taking the meter reading into account means that the system must rather be described by the conditional density matrix $\sigma_1(\,|m_k)$. See the text for further details.

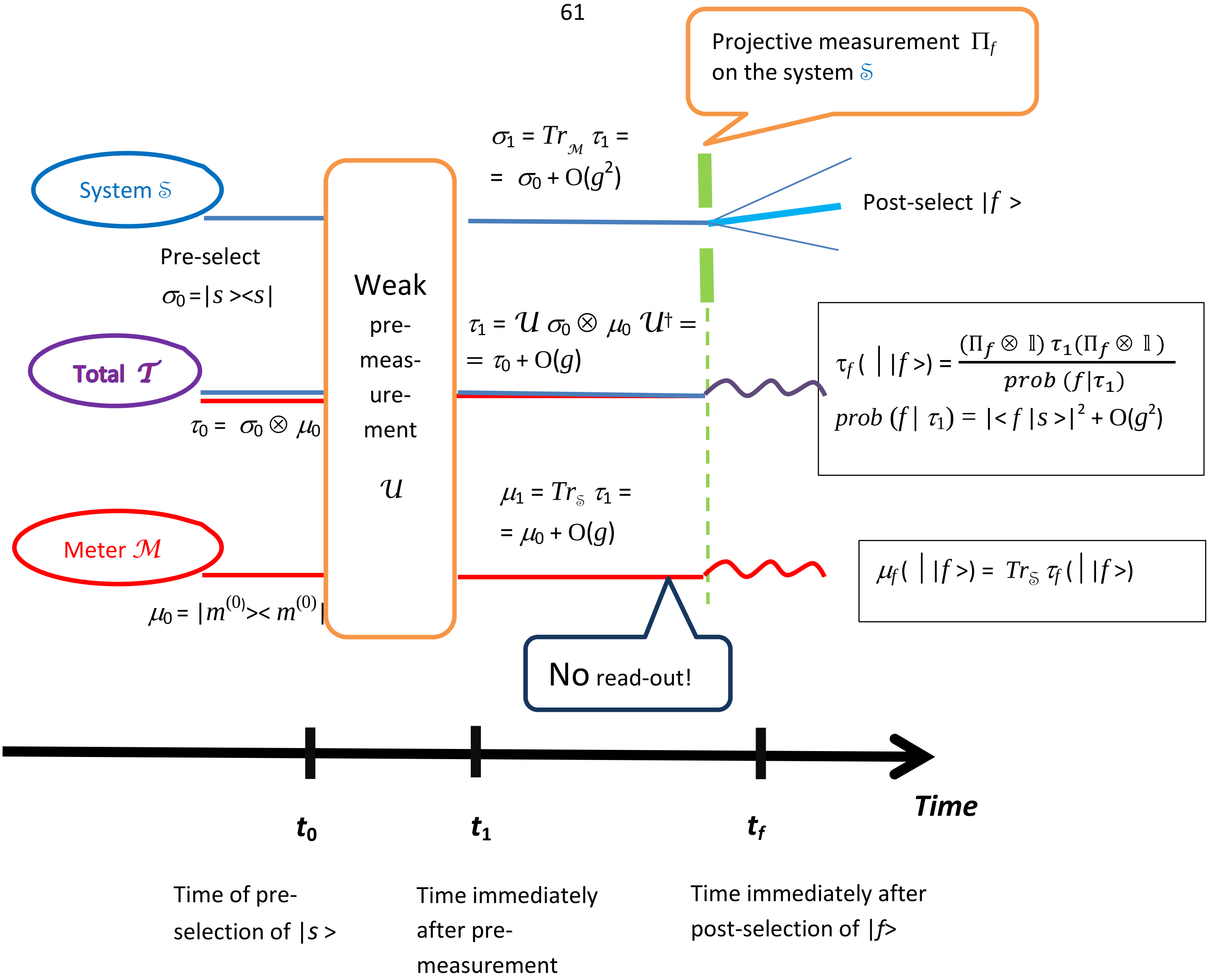

Figure 4. Schematic illustration of the weak measurement + post-selection protocol. The system is prepared ('pre-selected') in a pure state $|s\rangle$ which, through the weak (pre-) measurement, becomes entangled with the meter intitial state $|m^{(0)}\rangle$ but is otherwise left almost undisturbed; the deviation from the pre-selected state is at most second order in the weak measurement strength $g$. Without any read-out of the meter, the system is then subjected to a projective measurement of some observable $F$, and only one particular eigenstate $|f\rangle$ of this observable is chosen ('post-selection'). Meter observables after this post-selection are calculated from the meter density matrix $\mu_f(\ |\ |f\rangle)$. They can be expressed in terms of the weak value $S_w = \frac{\langle f\,|S|\,s\rangle}{\langle f\,|s\rangle}$ for the system observable $S$. It is assumed that there is no other time evolution but the measurements. See the text for more details.

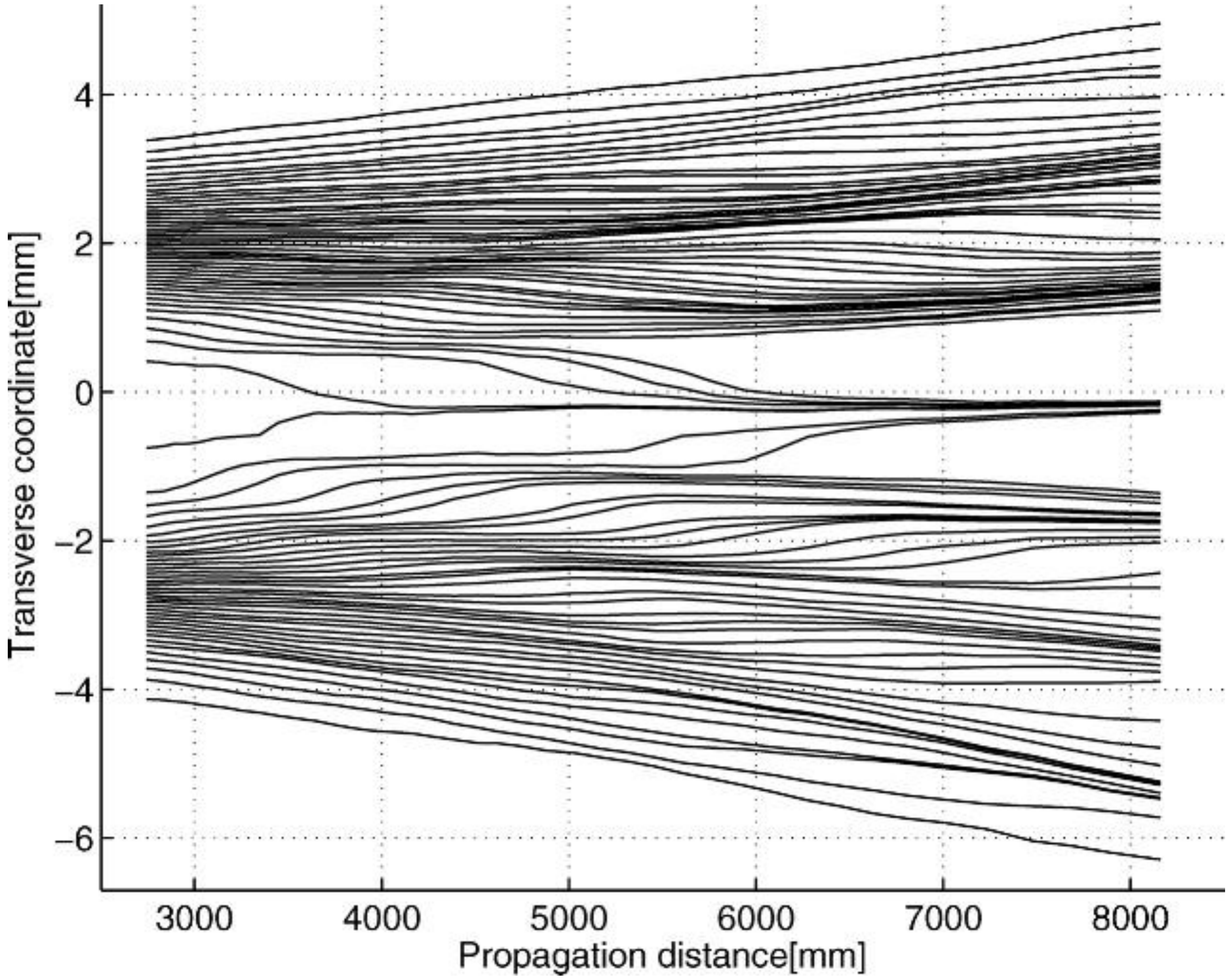

Figure 5. Trajectories for photons in a double-slit set-up as reviewed in section VII.D. The results are from the experiment by Kocsis et al [37]. The density of the trajectories at a particular location ('screen') is proportional to the number of photons hitting that screen, so the usual interference pattern may also be discerned. (Reproduced with permission from the author)

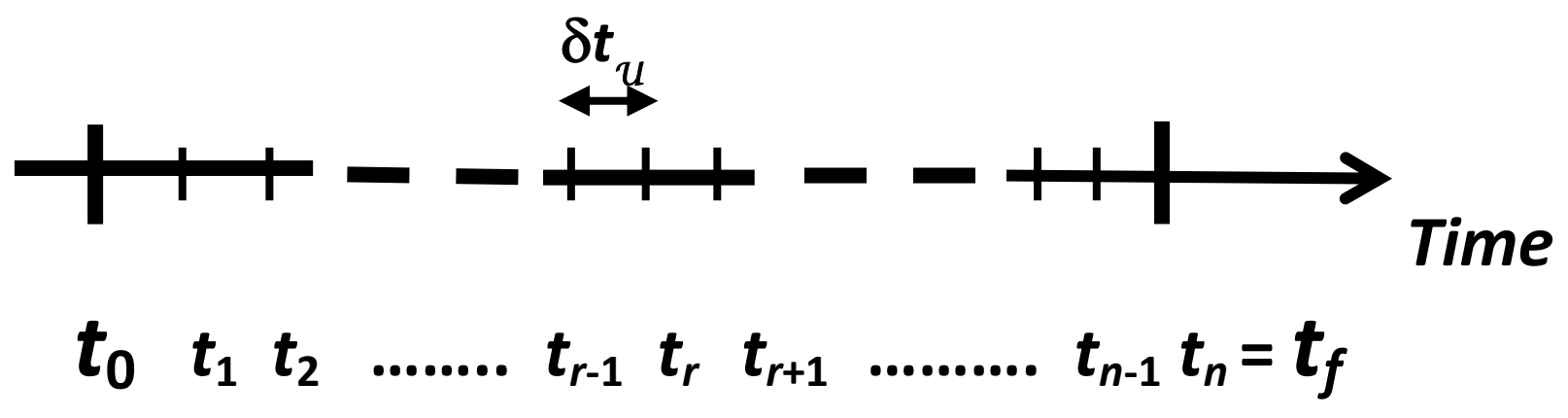

Fig 6. Illustration of the subdivision of the time interval between an initial time $t_0$ and a final time $t_f$ in $n$ equal steps of duration $\delta t_u$ , used in discussing the transition to continuous measurements in Appendix 2.

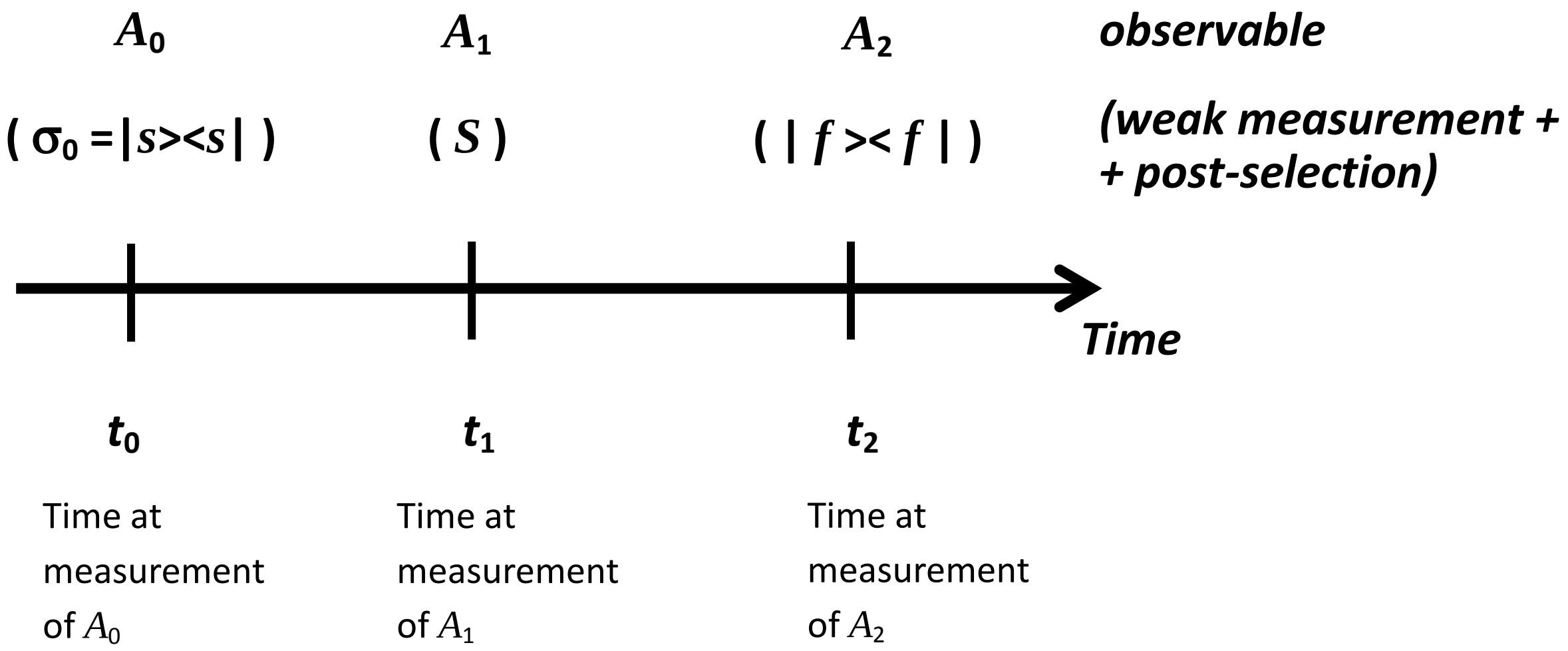

Figure 7. An illustration of the set-up behind the Leggett-Garg inequality. The different observables $A_0$, $A_1$ and $A_2$, are measured at consecutive times $t_0$, $t_1$ and $t_2$ (with no intrinsic time evolution in between). The symbols in brackets refer to the particular realizations of these observables in the weak measurement case as treated in the text.

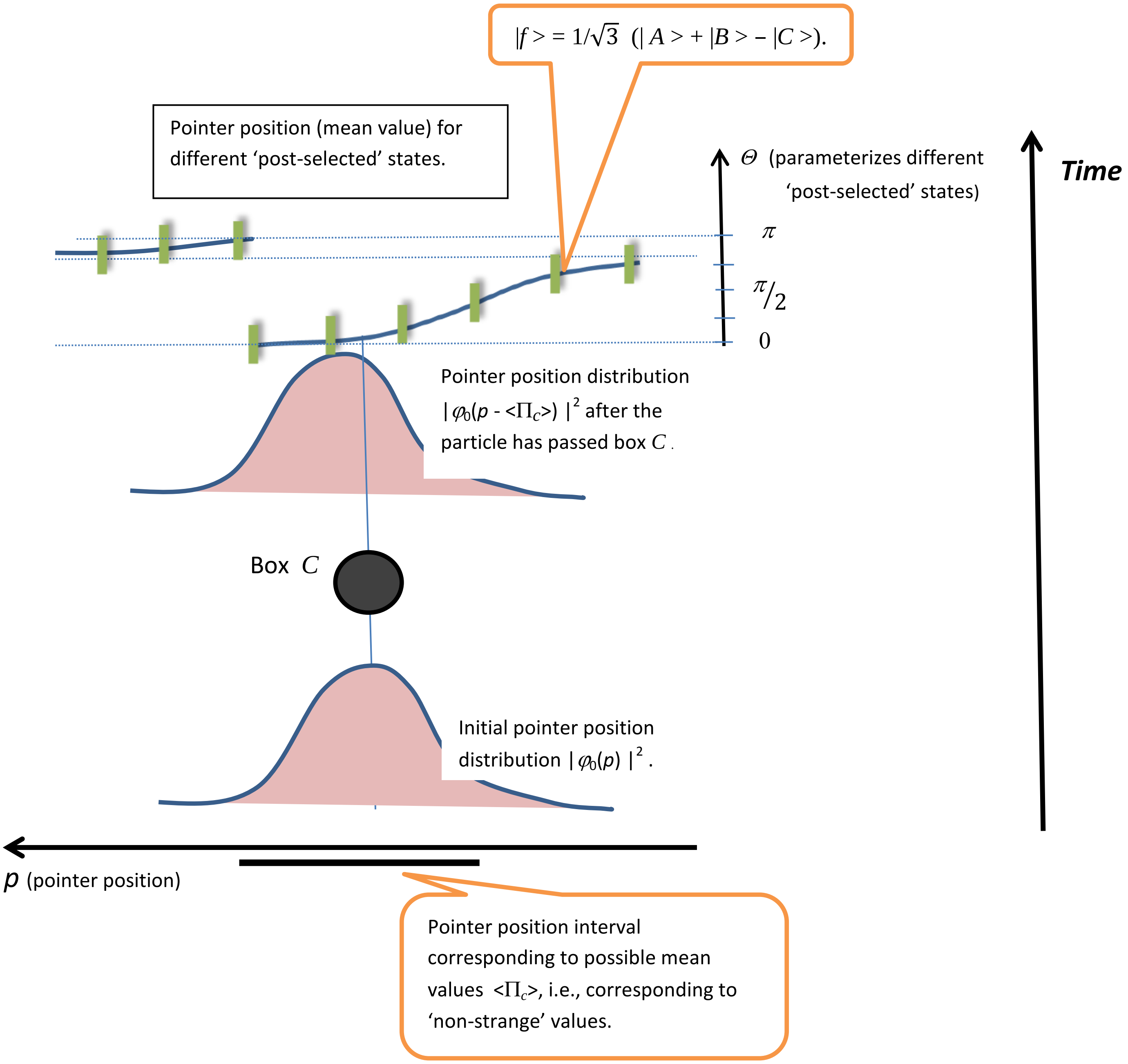

Figure 8. Schematic illustration of the analysis of the Three-Box Paradox, with time running up-ward. The 'post-selected' states, drawn as green 'slots' in the figure, are parameterized by an angle $\Theta$ (see text). They effectively act as 'filter' for the pointer position (*i.e.,* the transverse momentum of the particle), which is directly related to the weak value of the number operator $\Pi_C$ . (The slots mark the mean value of the pointer position; what is not shown is the fact that – also after post-selection – the spread of the pointer position distribution is approximately as large as in the initial state.) The fact that there can be very large values of the mean pointer position, even large negative ones, is due to the spread of the pointer position distribution after the particle has passed the box $C$ , *i.e.*, immediately before the post-selection. Note, however, that these 'strange values' occur mostly in the tails of the distribution, *i.e.*, with very low probability.